

\global\newcount\meqno
\def\eqn#1#2{\xdef#1{(\secsym\the\meqno)}
\global\advance\meqno by1$$#2\eqno#1$$}
%
\global\newcount\refno
\def\ref#1{\xdef#1{[\the\refno]}
\global\advance\refno by1#1}
\global\refno = 1
\vsize=7.5in
\hsize=5in
\magnification=1200
\tolerance 10000
%
%
%
%
\def\dofig{0}
%
%
%
\font\sevenrm=cmr7
\hyphenation{non-over-lapping}

\def\calr{ {\cal R}}

\def\eps{\epsilon}

\def\gsim{ \,\, \vcenter{\hbox{$\buildrel{\displaystyle >}\over\sim$}}
 \,\,}

\def\hk{\hat k}

\def\intz{\int_0^{\infty}dz~}

\def\lb{\lambda_{\beta}}

\def\lbk{\lambda_{\beta,k}}

\def\lsim{ \,\, \vcenter{\hbox{$\buildrel{\displaystyle <}\over\sim$}}
 \,\,}
\def\mapright#1{\smash{\mathop{\longrightarrow}\limits^{#1}}}
\def\mb{\mu_{\beta}}
\def\mbk{\mu_{\beta,k}^2}
\def\mbb{\mu_{\beta}^2}
\def\mk{\mu_{\beta,k}}

\def\ol{\bar}
\def\parmu{\partial_{\mu}}

\def\s#1{{\bf#1}}

\def\sqk{\sqrt{k^2+\tilde\mu^2_R}}
\def\sqkb{\sqrt{k^2+\mbk}}

\def\sqv{\sqrt{k^2+V''(\Phi)}}

\def\sumn{\sum_{n=-\infty}^{\infty}}

\def\tlambda{\tilde\lambda}
\def\tlk{\tilde\lambda_{\beta,k}}
\def\tmk{\tilde\mu^2_{\beta,k}}
\def\tmu{\tilde\mu}
\def\ubk{U_{\beta,k}(\Phi)}

\def\uhk{\hat U_{\hk}(\Phi)}

\def\pmb#1{\setbox0=\hbox{$#1$}%
\kern-.025em\copy0\kern-\wd0
\kern.05em\copy0\kern-\wd0
\kern-.025em\raise.0433em\box0 }
%
%
\ifnum\dofig=1
\input epsf
\fi
%
%
\baselineskip=0.1cm
\medskip
\nobreak
\medskip

\baselineskip 12pt plus 1pt minus 1pt
\vskip 2in
\centerline{\bf RENORMALIZATION GROUP APPROACH TO FIELD THEORY}
\vskip 12pt
\centerline{{\bf AT FINITE TEMPERATURE}}
\vskip 24pt
\centerline{Sen-Ben Liao and Michael Strickland}
\vskip 12pt
\centerline{\it Department of Physics}
\centerline{\it Duke University }
\centerline{\it Durham, North Carolina\ \ 27708\ \ \ U.S.A.}

\vskip 0.75in

\baselineskip 12pt plus 2pt minus 2pt
\centerline{{\bf ABSTRACT}}
\medskip
Scalar field theory at finite temperature is investigated
via an improved renormalization group prescription which provides an
effective resummation over all possible non-overlapping higher loop graphs.
Explicit analyses for the $\lambda\phi^4$ theory are performed in $d=4$
Euclidean space for both low and high temperature limits.
We generate a set of coupled equations
for the mass parameter and the coupling constant from the
renormalization group flow equation.
Dimensional reduction and symmetry restoration
are also explored with our improved approach.

\vskip 1.5in
\centerline{Submitted to {\it Phys. Rev. D}}
\vfill
\noindent DUKE-TH-94-81\hfill January 1995
\eject

\centerline{\bf I. INTRODUCTION}
\medskip
\nobreak
\xdef\secsym{1.}\global\meqno = 1
\medskip
\nobreak

In recent years, intensive efforts have been devoted to
quantum field theory at finite temperature, a subject with wide
applications in areas such as the evolution of the early universe and its
cosmological consequences \ref\linde, the deconfinement phenomena and
formation of the quark-gluon plasma \ref\muller, and the critical behavior of
condensed matter systems near the phase transition. To investigate these
issues, one
generally utilizes the finite-temperature effective potential approach
in the spirit of the perturbative loop expansion. However, in the high
temperature limit, perturbation theory becomes unreliable \ref\weinberg\
since the presence of infrared (IR) divergences may
destroy the correspondence between the expansions of loops and the
coupling constant. Certain higher loop contributions such as the
``daisy'' and ``superdaisy'' diagrams that contribute to the same
order in the coupling constant must also be incorporated \ref\jackiw\
for computing the critical transition temperature $T_c$ and determining
the nature of the
phase transition. In gauge theories, it has been shown that the ``hot thermal
loops'' need to be resummed in order to obtain a gauge independent gluon
damping rate \ref\bp.

However, since resumming the multi-loop contributions to
arbitrary orders proves to be nontrivial,
various methods have been proposed to carry out the resummation:
In \jackiw, the use of a gap equation was first discussed; in
\ref\elmfors\ a renormalization
group (RG) with $T$, the temperature, as the flow parameter was used; and
in \ref\sypi\ a self-consistent Hartree-Fock formalism is presented.
An ``environmentally friendly'' renormalization prescription for interpolating
effective finite-temperature theories in different regimes has also
been proposed \ref\oconnor.
Although all of these techniques yield the same finite-temperature
effective propagators in the leading order, they differ in the
subleading correction which can affect the nature of
the phase transition.

In this paper, the methodology adopted for investigating the scalar field
theory at finite temperature
is based on the use of RG constructed from the Wilson-Kadanoff \ref\wils\
blocking transformation in the Euclidean formalism. Unlike \elmfors,
our RG arises from the arbitrariness of the internal blocking scale $k$,
and not the external temperature parameter $T$.
The formulation not only takes into consideration the dominant
higher loop diagrams without resorting to the complicated analytical
order by order resummation,
it also characterizes the flow pattern of the theory for arbitrary $T$
as well as the momentum scale $k$ which is chosen to be zero in the
methods described above. Our approach is analogous to the series of works
by Tetradis and Wetterich \ref\wetterich. However, instead of using
a smooth momentum smearing function which leads to an integro-differential
RG equation, our sharp momentum cut-off \ref\lpt\ yields a RG flow equation
which takes on the form of non-linear partial differential equation.
Besides the advantage in performing numerical computation, it also offers
a more lucid physical interpretations.

Consider the following scalar lagrangian:
\eqn\slan{ {\cal L}={1\over 2}(\parmu\phi)^2+V(\phi).}
At finite temperature, the RG improved blocked potential $\ubk$ associated
with the blocked field $\Phi(x)$ is characterized by the following
differential RG equation \lpt\ :
\eqn\rgfto{k{{\partial U_{\beta,k}}\over {\partial k}}
=-{k^3\over4\pi^2}\sqrt{k^2+U''_{\beta,k}}
-T{k^3\over2\pi^2}{\rm ln}\Bigl[1-e^{-\beta\sqrt{k^2+U''_{\beta,k}}}\Bigr].}
The above equation is obtained by first computing the finite-temperature
blocked potential to the one-loop order followed by a RG improvement
to take into account all possible non-overlapping
daisy, superdaisy and higher-loop diagrams which strongly
modify theory in the high $T$ and small $k$ limits.
We also see that in this RG construction the contributions
of a particular mode which has been integrated out are naturally
retained for the
integration of the next, and therefore the interactions
among the modes are properly taken into account. In addition, during the
course of mode elimination,
irrelevant operators defined with respect to the ultraviolet (UV) fixed point
continue to be generated and their effects
incorporated as well by \rgfto \ref\llp. Furthermore, our flow equation also
offers insights into the existence of high temperature dimensional
reduction as well as the restoration of symmetry accompanied by the
disappearance of imaginary contribution to $\ubk$, as we shall see later.
To describe the full theory, however, one also needs to consider
the effects of wavefunction renormalization
constant $Z_{\beta,k}(\Phi)$ as well as the higher-order derivative terms
in the blocked lagrangian. Nevertheless, the use of \rgfto\
is justified in the IR limit of the four-dimensional theories since
the other contributions are only of higher order.

The organization of the paper is as follows: In Sec. II
the formalism of the blocking transformation at finite temperature is briefly
reviewed using the scalar $\lambda\phi^4$ theory as illustration. The
finite temperature RG flow equation \rgfto\ with underlying $O(3)$ symmetry
is constructed and
compared with the simple one-loop independent mode approximation (IMA).
In Sec. III we concentrate on the low $T$ regime where thermal effects
are negligible and compare the $O(3)$ thermal blocked potential $\ubk$ with
the $O(4)$ symmetric $U_k(\Phi)$ at $T=0$. The high $T$ limit of
the theory is investigated in Section IV. The criterion for dimensional
reduction is explicitly deduced from \rgfto. In addition, the behaviors
of the scale-dependent thermal mass and coupling constant $\mu_{\beta,k}^2$
and $\lambda_{\beta,k}$, are studied with a set of coupled equations.
It is shown that $\lambda_{\beta,k}$ decreases with increasing $T$ and
approaches a constant in this regime. The complementary relationship
between the
internal scale $k$ and the external parameter $T$ is also demonstrated.
The phenomenon of spontaneous symmetry breaking at $T=0$ and symmetry
restoration above $T_c$ is discussed in Sec. VI. The phase boundary
is determined from \rgfto\ by requiring the imaginary part of $\ubk$
to vanish at $T_c$. The value obtained in this manner is compared with
that obtained in the one-loop approximation without the imaginary sector.
Sec. VI is reserved for summary and discussions. We collect in the
Appendix the details of extracting the leading order thermal contributions
from the integrals encountered in Sec. IV.

\medskip
\medskip
\centerline{\bf II. FINITE-TEMPERATURE RENORMALIZATION GROUP}
\medskip
\nobreak
\xdef\secsym{2.}\global\meqno = 1
\medskip
\nobreak

In this section, we briefly review the finite temperature RG formalism
used in investigating the thermal behavior of the theory.
The concept of the RG is based on the notion that in certain
physical processes only a particular range of modes in the momentum
decomposition of the field $\phi(x)$ will be relevant, and it is often
desirable to eliminate the irrelevant modes to which the
physics is insensitive. The reduction of degrees of freedom is then
compensated by a readjustment of the parameters in the coupling constant
space of the lagrangian. Therefore,
instead of using the original field variable
\eqn\ppx{\phi(x)={1\over\beta}\sum_{n=-\infty}^\infty\int_{\s p}
e^{-i(\omega_n\tau-\s p\cdot\s x)}\phi(\omega_n,\s p), \qquad
\int_{\s p}=\int {d^3\s p\over(2\pi)^3},}
where $\beta^{-1}=T$ and $\omega_n={2\pi n}/{\beta}$ denotes the Matsubara
frequency,
%
%
we define the coarse-grained blocked field as \lpt:
\eqn\blctk{\Phi(\s x)=T\int_0^\beta dy^0\int d^3\s y
\rho_k(\s x-\s y)\phi(y)}
via an $O(3)$ invariant smearing function $\rho_k(\s x)$. Since the
low energy physics
is unaffected by the modes above the ``blocking scale'' $k$, we shall
for simplicity choose
\eqn\smfrk{\rho_k(\s x)=\int_{|\s p| < k}{d^3\s p\over(2\pi)^3}
e^{i\s p\cdot\s x}, }
or $\rho_k(\s p)=\Theta(k-|\s p|)$. In other words,
$k$ acts as an upper cut-off for the modes which are to be retained.
Notice that in this formulation, the $\delta$-
function associated with the imaginary-time variable $\tau$ is given by
\eqn\erfd{ \delta(\tau_x-\tau_y)={1\over\beta}\sumn e^{-i\omega_n(\tau_x
-\tau_y)}.}

Since the wavefunction renormalization constant $Z_k(\Phi)$ yields only
a minute correction to the anomalous dimension, we shall for simplicity
set $Z_k(\Phi)$ to be unity. In addition, by neglecting the higher order
field derivative terms, one may simply choose the static limit
$\Phi(\s x)=\Phi=const.$ for computing $\ubk$. Upon a simple Gaussian
integration, the
one-loop contribution to the finite-temperature blocked potential
$\tilde\ubk$ becomes
\eqn\uc{\eqalign{ \tilde U^{(1)}_{\beta,k}(\Phi) &=
{1\over2\beta}\sumn\int_{\s p}^{'}{\rm ln}
\Bigl[\omega_n^2+p^2+V''(\Phi)\Bigr]\cr
&={1\over 2\beta}\int_{\s p}^{'}
\Biggl\{\beta\sqrt{p^2+V''(\Phi)}+2{\rm ln}
\Bigl[1-e^{-\beta\sqrt{p^2+V''(\Phi)}}\Bigr]\Biggr\},}}
where the prime on the $\s p$
integral indicates an integration subject to the constraint $k \le {\s p}
\le\Lambda$, with $\Lambda$ chosen as the UV regulator for the
three-momentum integration.
For $V(\Phi)=\mu^2\Phi^2/2 +\lambda\Phi^4/4!$, by demanding the renormalized
parameters to satisfy
\eqn\hemoo{\cases{\eqalign{\tilde\mu^2_R&={\partial^2{\tilde U_{\beta,k}}
\over\partial\Phi^2}\Big\vert_{\Phi=\beta^{-1}=k=0}\cr
\tilde\lambda_R&={\partial^4{\tilde U_{\beta,k}}\over\partial\Phi^4}
\Big\vert_{\Phi=\beta^{-1}=k=0,} \cr}}}
the resulting $\tilde\ubk$ takes on the form:
\eqn\upk{\eqalign{\tilde U_{\beta,k}(\Phi)&={\tilde\mu_R^2\over2}\Phi^2
\Bigl(1-{\tilde\lambda_R\over64\pi^2}\Bigr)
+{\tilde\lambda_R\over4!}\Phi^4\Bigl(1-{9\tilde\lambda_R\over64\pi^2}\Bigr)\cr
&+{1\over32\pi^2}\Biggl\{-k\Bigl(2k^2+\tilde\mu_R^2+{1\over2}\tilde\lambda_R
\Phi^2\Bigr)\Bigl({k^2+\tilde\mu_R^2+{1\over2}\tilde\lambda_R\Phi^2}
\Bigr)^{1/2}\cr
&
+\Bigl(\tilde\mu_R^2+{1\over2}\tilde\lambda_R\Phi^2\Bigr)^2{\rm ln}
\Bigl[{k+{\sqrt {k^2+\tmu_R^2+\tlambda_R\Phi^2/2}}\over\tmu_R}\Bigr]\Biggr\}\cr
&+{1\over2\pi^2\beta}\int_k^{\Lambda}dpp^2{\rm ln}
\Bigl[1-e^{-\beta\sqrt{p^2+\tmu_R^2+\tlambda_R\Phi^2/2}}\Bigr],}}
which in the limits of vanishing $\beta^{-1}$ and $k$, reproduces the
usual effective potential \ref\coleman. Therefore, one sees that
consistency with the $O(4)$ invariant theory in the $T=0$
limit requires a complete summation over the Matsubara
frequencies $\omega_n=2\pi n/{\beta}$ for all $n$, i.e., $\rho_k(\s p)$
must be independent of $\omega_n$. It is also evident
from \hemoo\ that temperature-independent subtractions alone are sufficient
to remove the cut-off dependence.

Differentiating \uc\ with respect to the arbitrary scale $k$ leads to
\eqn\rgfti{k{{\partial {\tilde U_{\beta,k}}}\over {\partial k}}
=-{k^3\over4\pi^2}\sqrt{k^2+V''(\Phi)}
-T{k^3\over2\pi^2}{\rm ln}\Bigl[1-e^{-\beta\sqrt{k^2+V''(\Phi)}}\Bigr].}
This RG equation is obtained in a manner in which
each individual mode is integrated out independently by neglecting
the systematic feedbacks from the high modes to the lower ones during
the elimination. The parameters for this independent-mode approximation
(IMA) scheme are denoted by a tilde.

On the other hand, instead of integrating out all the modes from $\Lambda$ to
$k$ all at once, one may first divide the integration volume into a large
number of thin shells of small thickness $\Delta k$. By lowering the
cut-off infinitesimally from $\Lambda\to \Lambda-\Delta k$ until
$\Lambda=k$ is reached, we arrive at the following
RG equation governing the flow pattern of the {\it improved}
finite-temperature blocked potential $\ubk$:
\eqn\rgft{k{{\partial U_{\beta,k}}\over {\partial k}}
=-{k^3\over4\pi^2}\sqrt{k^2+U''_{\beta,k}}
-T{k^3\over2\pi^2}{\rm ln}\Bigl[1-e^{-\beta\sqrt{k^2+U''_{\beta,k}}}\Bigr].}
This non-linear partial differential equation establishes a smooth connection
between the small- and large-distance physics at finite temperature.
Moreover, \rgft\ systematically
incorporates the contribution of a particular mode for the elimination
of the next. In fact, \rgft\ takes into account all possible non-overlapping
daisy, superdaisy and higher-loop diagrams which significantly modify the
theory at high $T$ and small $k$ limits. The summation over
the multi-loop graphs from the basic one-loop structure is depicted in
Fig. 1. In other words, what we have here is a ``physical'' one-loop
diagram characterized by the dressed vertices.
One may certainly improve \rgft\ by including higher ``physical'' loop
contributions. However, we argue that since each loop integration is
multiplied by a factor
$\kappa={\Delta k}/k$, a term of order $m$ in loops will be suppressed by
$\kappa^m$ in the limit $\Delta k\to 0$. Therefore, the use of
\rgft\ is justified.
What we have neglected in \rgft\ are the overlapping higher loop diagrams.
However, as shown in \jackiw, if one considers the $N$-component theory
and takes the limit $N\to\infty$, these overlapping graphs will be
suppressed by an extra factor $N$ compared with the corresponding
non-overlapping
contributions at the same loop order. The effects of these overlapping
graphs for the one-component case is currently being explored.

Eq. \rgft\ implies the following RG flow equation for
the effective scale-dependent thermal parameters
$\mu^2_{\beta,k}=U''_{\beta,k}(0)$, $\lambda_{\beta,k}=U^{(4)}_{\beta,k}(0)$
and $g_{\beta,k}=U^{(6)}_{\beta,k}(0)$:
\eqn\efmass{ k{{\partial\mu_{\beta,k}^2}\over\partial k}=
-{\lbk\over 8\pi^2}{k^3\over\sqrt{k^2+\mbk}}~{\rm coth}\Bigl(
{\beta\sqrt{k^2+\mbk}\over 2}\Bigr),}
and
\eqn\efcou{\eqalign{ k{\partial\lbk\over\partial k}&={3\lbk^2\over 16\pi^2}
{k^3\over{k^2+\mbk}}\Biggl\{{1\over\sqrt{k^2+\mbk}}~{\rm coth}\Bigl({\beta
\sqrt{k^2+\mbk}\over 2}\Bigr)+{e^{\beta\sqrt{k^2+\mbk}}\over
{T\bigl(e^{\beta\sqrt{k^2+\mbk}}-1\bigr)^2}}\Biggr\} \cr
&
-{3g_{\beta,k}\over 8\pi^2}{k^3\over\sqrt{k^2+\mbk}}\Bigl[1+{2\over
{3\bigl(e^{\beta\sqrt{k^2+\mbk}}-1\bigr)}}\Bigr].}}
Notice the contribution to the flow of $\lambda_{\beta,k}$ from the
sixth-order coupling constant $g_{\beta,k}$. The presence of such higher
order corrections is the natural consequence of blocking transformation
through which the number of effective modes is reduced at
the expense of generating a more complicated effective action.

\ifnum\dofig=1
\bigskip

\centerline{\epsfbox{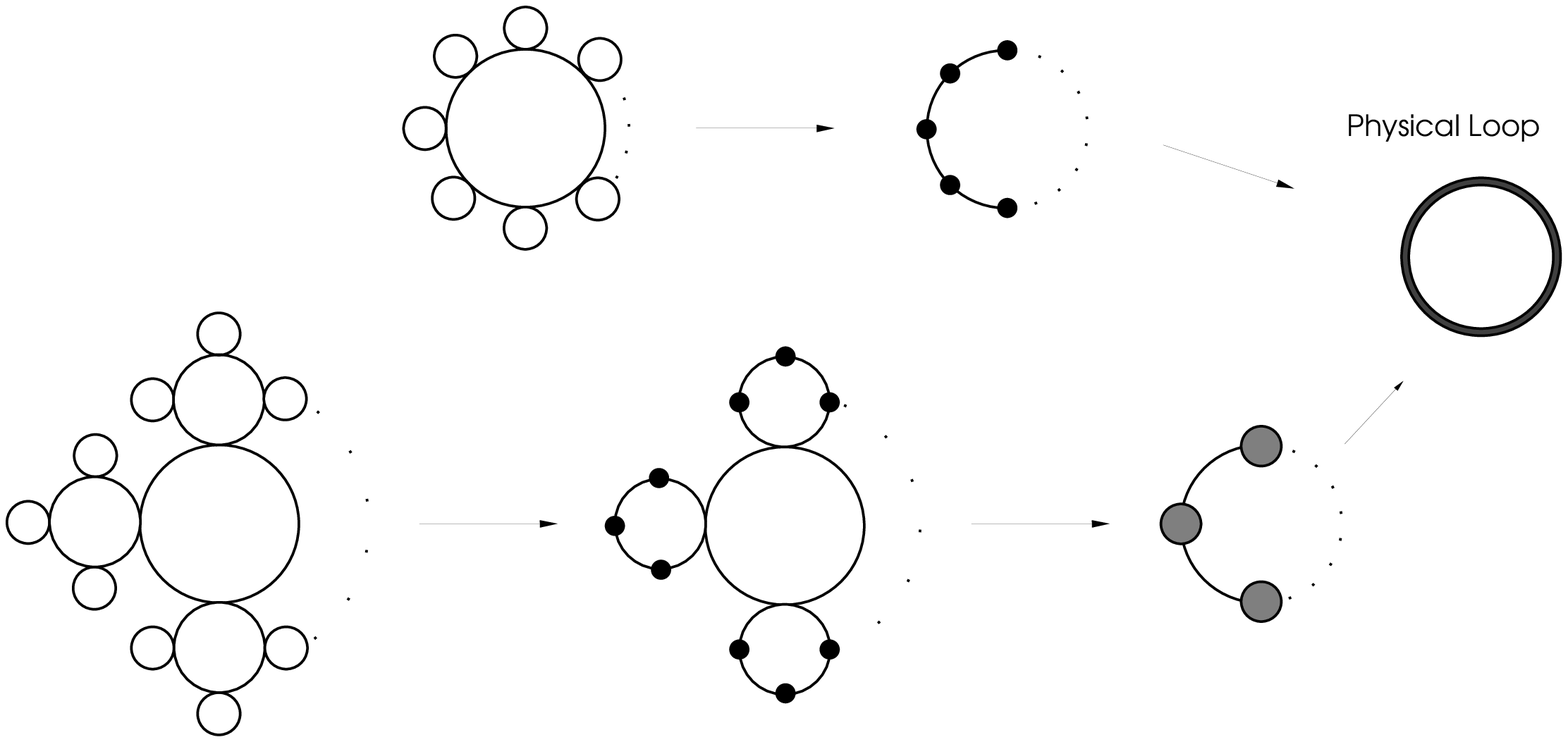}}
\bigskip
{\narrower
{\sevenrm
{\baselineskip=8pt
\itemitem{Figure 1.}
Diagrammatic representation of loop resummation.  Contributions from all
non-overlapping graphs are included in our physical loop.
\bigskip
}}}
\fi

While $\tilde U_{\beta,k}(\Phi)$ associated with the IMA
is solved analytically in \upk\ apart from the integral containing thermal
effects, the RG improved $U_{\beta,k}(\Phi)$ can only be obtained by
solving \rgft\ numerically.
In Figs. 2 and 3, the evolution of $U_{\beta,k}(\Phi)$ and
$\tilde U_{\beta,k}(\Phi)$ for $T=0$ and $T\ne 0$ are traced.
We notice that the
two potentials agree reasonably well for $k$ near the cut-off $\Lambda$
and begin to deviate as $k$ is lowered. In the deep IR limit where
$k=0$, appreciable difference between $U_{\beta,k}(\Phi)$ and
$\tilde U_{\beta,k}(\Phi)$ is seen. Such a difference can be
understood by noting that
\eqn\dife{\eqalign{k{\partial\over\partial k}\Bigl(U_{\beta,k}(\Phi)
-{\tilde U}_{\beta,k}(\Phi)\Bigr)&=-{k^3\over{4\pi^2}}\Biggl\{
\sqrt{k^2+U''_{\beta,k}(\Phi)}-\sqrt{k^2+V''(\Phi)}\Biggr\} \cr
&\qquad\quad
-{Tk^3\over{2\pi^2}}{\rm ln}\Biggl[{{1-e^{-\beta\sqrt{k^2+U''_{\beta,k}
(\Phi)}}}\over{1-e^{-\beta\sqrt{k^2+V''(\Phi)}}}}\Biggr] \cr
&
=-{k^3\over{4\pi^2}}\sqrt{k^2+V''(\Phi)}\Biggl\{\Bigl[1+{{U_{\beta,k}''(\Phi)
-V''(\Phi)}\over{k^2+V''(\Phi)}}\Bigr]^{1/2}-1\Biggr\}\cr
&
-{Tk^3\over{2\pi^2}}{\rm ln}\Biggl[1+{{{\rm exp}\bigl\{-\beta\bigl(
\sqrt{k^2+V''(\Phi)}-\sqrt{k^2+U''_{\beta,k}(\Phi)}\bigr)\bigr\}}\over
{1-{\rm exp}\bigl\{-\beta\sqrt{k^2+V''(\Phi)}\bigr\}}}\Biggr] \cr
&
=-{k^3\over{8\pi^2}}{{\bigl(U_{\beta,k}''(\Phi)-V''(\Phi)\bigr)}\over\sqv}
\Biggl[1-{1\over 4}\Bigl({{U_{\beta,k}''(\Phi)-V''(\Phi)}
\over{k^2+V''(\Phi)}}\Bigr)\Biggr]\cr
&\qquad
\times\Biggl\{ 1-{2T\over\sqv}\Bigl(e^{\beta\sqv}-1\Bigr)^{-1}\Biggr\}
+\cdots.}}
Thus we see that higher loop corrections will
continue to pile up as $k$ is lowered or $T$ is raised.
By comparing Fig.2
and Fig. 3, we see that the inclusion of the thermal effects improves the
agreement between the RG and the IMA results. In Sec. IV we shall
see that the high $T$ behavior of the theory is strongly modified by
the higher loop contributions.

\ifnum\dofig=1
\bigskip

\centerline{\epsfbox{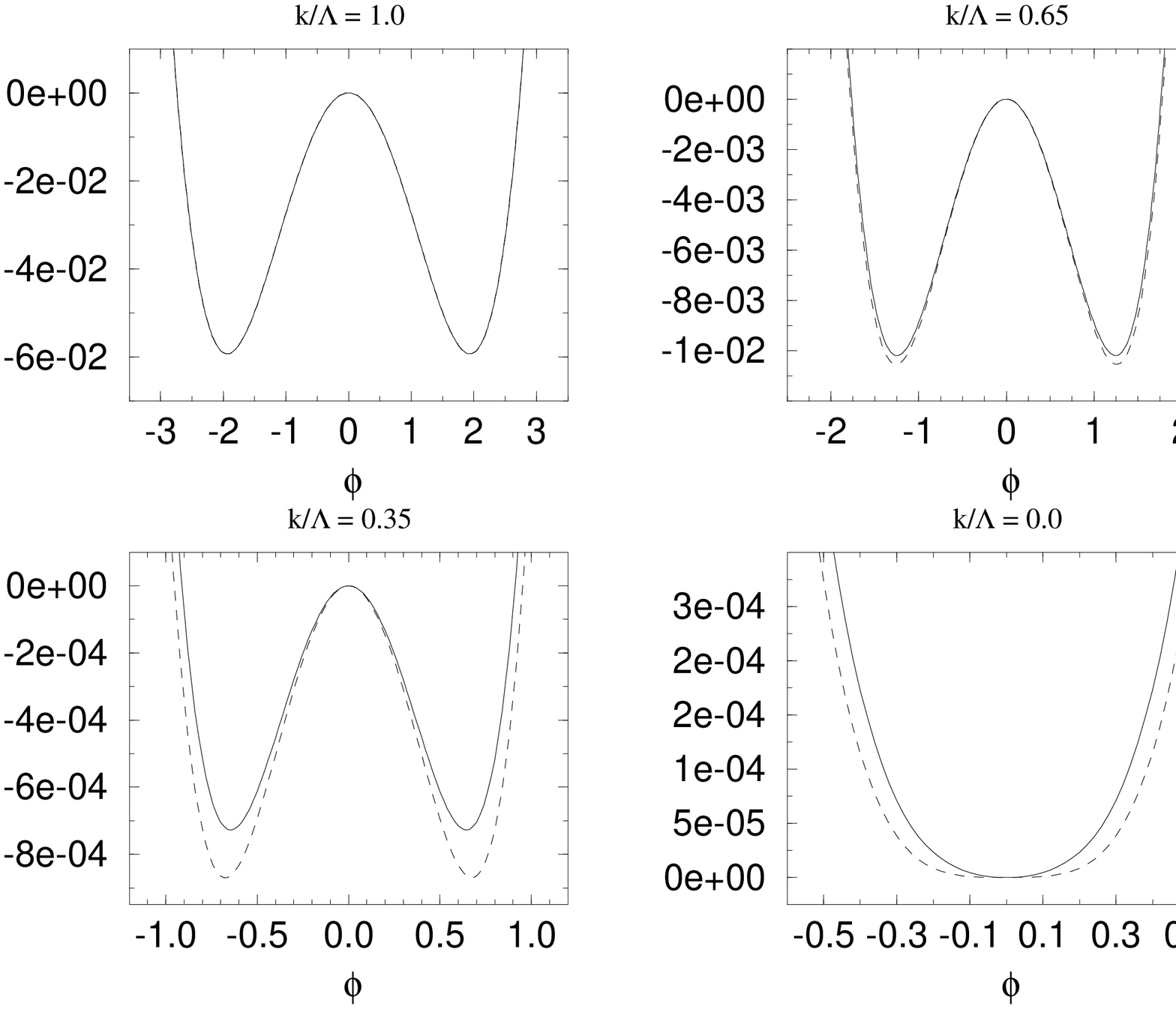}}
\bigskip
{\narrower
{\sevenrm
{\baselineskip=7pt
\itemitem{Figure 2.}
Flow pattern of the blocked potential for
$\scriptstyle T=0$ with $\scriptstyle \tilde\mu_R^2=10^{-4}$, $\scriptstyle
\tilde\lambda_R=0.1$ and $\scriptstyle \Lambda=10$.
Solid and dashed lines represent the RG and IMA results, respectively.
\bigskip
}}}
\fi

\ifnum\dofig=1
\bigskip

\centerline{\epsfbox{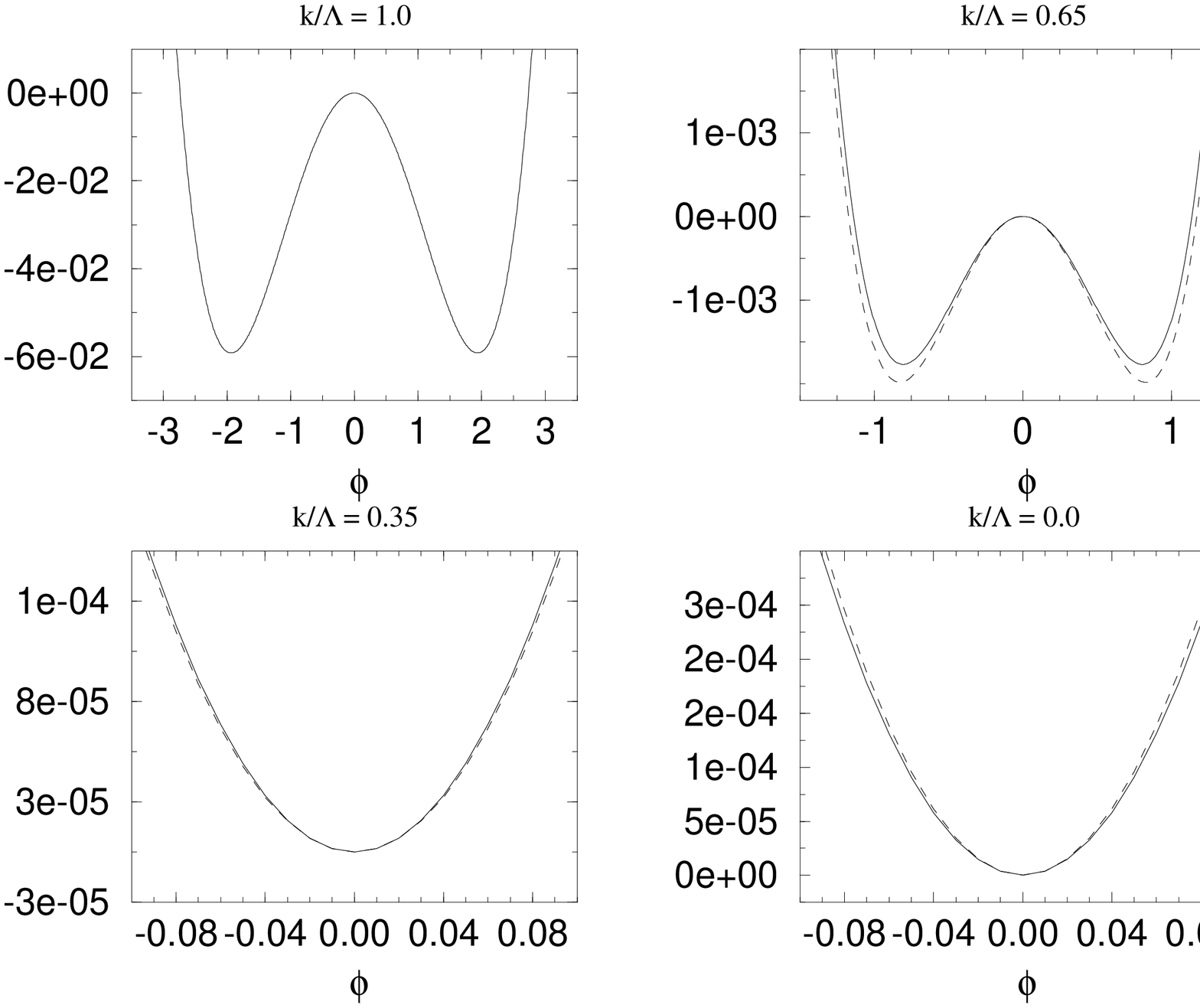}}
\bigskip
{\narrower
{\sevenrm
{\baselineskip=7pt
\itemitem{Figure 3.}
Flow pattern of the blocked potential for
$\scriptstyle T=5$ with $\scriptstyle \tilde\mu_R^2=10^{-4}$, $\scriptstyle
\tilde\lambda_R=0.1$ and $\scriptstyle \Lambda=10$.
Solid and dashed lines represent the RG and IMA results, respectively.
\bigskip
}}}
\fi

In Fig. 4 the evolutions of $\mu^2_{\beta,k}$ and $\lambda_{\beta,k}$
with $k$ for three different values of $T$ are depicted. While
$\mu^2_{\beta,k}$
increases with $T$,
$\lambda_{\beta,k}$ is shown to decrease with rising $T$. On the other hand,
when the role of $k$ is considered, its influences on $\mu^2_{\beta,k}$ and
$\lambda_{\beta,k}$ are completely opposite to that of $T$. One therefore
observes an interesting competition between the external parameter $T$
and the internal momentum scale parameter $k$ from which RG is defined.
By external parameter we mean a ``physical'' quantity which has direct
physical impact
on the system and can be measured experimentally. On the other hand,
an internal scale is a ``fictitious'' scale which we choose to characterize
the theory. In this case, $k$ is chosen to provide a separation between
the high and the low modes. However, the precise definition of ``high'' and
``low'' modes is dependent on the energy regime under investigation.
A more complete discussion concerning the behaviors of the thermal parameters
for different $k$ will appear in the later sections.

\ifnum\dofig=1
\bigskip

\centerline{\epsfbox{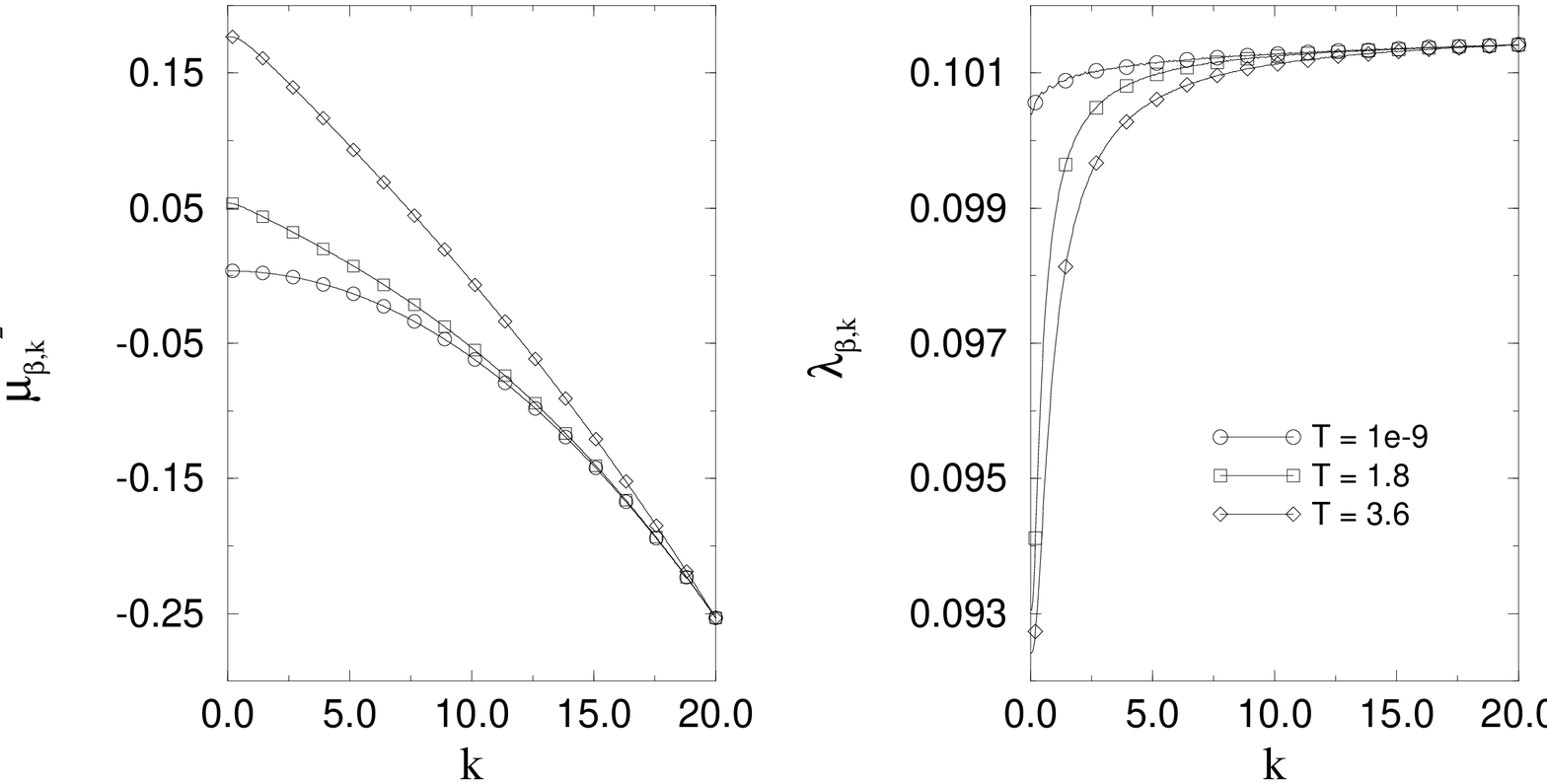}}
\bigskip
{\narrower
{\sevenrm
{\baselineskip=7pt
\itemitem{Figure 4.}
Evolution of $\scriptstyle \mu_{\beta,k}^2$ and
$\scriptstyle \lambda_{\beta,k}$ as a function of $\scriptstyle k$ for
various values of $\scriptstyle T$ using
$\scriptstyle \tilde\mu_R^2=10^{-4}$, $\scriptstyle \tilde\lambda_R=0.1$,
and $\scriptstyle \Lambda = 20$.
\bigskip
}}}
\fi

\bigskip
\medskip
\centerline{\bf III. LOW TEMPERATURE LIMIT}
\medskip
\nobreak
\xdef\secsym{3.}\global\meqno = 1
\medskip
\nobreak

We first consider the low temperature limit
$T<<\sqrt{k^2+U''_{\beta,k}}$ where, upon dropping the subscript $\beta$,
\rgft\ is reduced to
\eqn\rgftl{k{{\partial U_k}\over {\partial k}}=
-{k^3\over4\pi^2}\sqrt{k^2+U''_k}.}
This ``zero'' temperature limit can be compared with the RG flow equation
associated with $\hat U_{\hk}(\Phi)$, the blocked potential for $d=4$
derived in an $O(4)$ symmetric manner \ref\lp:
\eqn\rgcoe{\hk\partial_{\hk} {\hat U_k(\Phi)}=-{\hk^4\over16\pi^2}{\rm ln}
\Bigl({{\hk^2+\hat U_{\hk}''(\Phi)}\over {\hk^2+\hat U''_{\hk}(0)}}\Bigr)}
Note that quantities derived from the $O(4)$ symmetric $\hat U_{\hk}(\Phi)$
shall be distinguished with a caret, and double carets for the corresponding
IMA scheme.

Since the underlying symmetry of the finite temperature RG equation \rgft\
is $O(3)$, it is instructive to compare its low $T$ flow patterns
with \rgcoe\ for both RG improved and IMA
prescriptions. Eqs. \rgftl, \rgcoe\
along with \rgfti\ lead to the following RG coefficient functions
for the running mass parameter and the running coupling constants:
\eqn\hemo{\cases{\eqalign{\tilde\beta_2&=
k{{\partial\tilde\mu_k^2}\over\partial k}=-{\tlambda_R\over 8\pi^2}
{k^3\over{\sqrt{k^2+\tilde\mu_R^2}}}\qquad~~~~~~\bigl[{\rm O(3),~IMA}\bigr] \cr
\beta_2&=k{{\partial\mu_k^2}\over\partial k}=-{\lambda_k\over 8\pi^2}
{k^3\over\sqrt{k^2+\mu^2_k}}\qquad~~~~~~~\bigl[{\rm O(3), ~RG}\bigr]\cr
{\hat{\hat\beta}}_2&=
\hk{{\partial {\hat\mu_{\hk}^2}}\over\partial\hk}=-{\hat\lambda_R\over 16\pi^2}
{\hk^4\over{\hk^2+\hat \mu_R^2}}\qquad ~~~~~~~~\bigr[{\rm O(4), ~IMA}\bigr]\cr
\hat\beta_2&=
\hk{{\partial {\hat\mu_{\hk}^2}}\over\partial \hk}=-{\hat\lambda_{\hk}
\over 16\pi^2}{\hk^4\over{\hk^2+\hat \mu_{\hk}^2}}\qquad ~~~~~~~~\bigl[
{\rm O(4), ~RG}\bigr]\cr}}}
\eqn\hbet{\cases{\eqalign{\tilde\beta_4&=k{{\partial\tilde\lambda_k}
\over\partial k}=
{3\tlambda_R^2\over 16\pi^2}{k^3\over{
(k^2+\tilde\mu_R^2)}^{3/2}} \qquad\qquad\qquad\qquad~~~~\bigl[{\rm O(3), ~IMA}
\bigr]\cr
\beta_4&=k{\partial\lambda_k\over\partial k}=
{3\lambda_k^2\over 16\pi^2}{k^3\over{(k^2+\mu_k^2)}^{3/2}}
-{g_k\over 8\pi^2}{k^3\over\sqrt{k^2+\mu_k^2}} ~~~~~~\bigl[{\rm O(3), ~RG}
\bigr]\cr
\hat{\hat\beta}_4&=\hk{{\partial\hat\lambda_{\hk}} \over\partial \hk}=
{3{\hat\lambda_R}^2\over 16\pi^2}{\hk^4\over{(\hk^2
+\hat\mu^2_R)}^2}\qquad\qquad\qquad\qquad\qquad\bigl[{\rm O(4), ~IMA}\bigr]\cr
\hat\beta_4&=\hk{{\partial\hat\lambda_{\hk}} \over\partial \hk}=
{3{\hat\lambda_{\hk}}^2\over 16\pi^2}{\hk^4\over{(\hk^2+\hat\mu^2_{\hk})}^2}
-{{\hat g_{\hk}}\over 16\pi^2}{\hk^4\over{\hk^2
+\hat\mu_{\hk}^2}}\qquad~~~~\bigl[{\rm O(4), ~RG}\bigr]. \cr}}}
Notice that in the RG improved $O(3)$ and $O(4)$ schemes for
$\lambda_k$, the influences from $g_k$ is incorporated.
For $k^2 >> \mu_k^2$ and $\hk^2 >>\hat\mu_k^2$, one finds agreement in the
leading order behavior of $\beta_4$ and $\hat\beta_4$
for the running coupling constant, as can be seen from \hbet. However,
for the running mass parameter,
$\beta_2$ and $\hat\beta_2$ differ by a factor of two in this limit. Such
a discrepancy can easily be understood by noting the difference in the
symmetries of the underlying RG constructs, namely, $O(3)$ for the former
and $O(4)$ for the latter. The manner in which the original bare theory is
approached can be elucidated by examining the simple one-loop structure
of the bare mass parameter:
\eqn\barm{\eqalign{\mu_B^2 &=\tmu_R^2-{\tlambda_R\over 32\pi^2}\Bigl[
2\Lambda^2-\tmu_R^2{\rm ln}\bigl({2\Lambda^2\over\tmu_R^2}\bigr)+\tmu_R^2
\bigl(1-{\rm ln}2\bigr)\Bigr] \qquad\qquad :O(3) \cr
&
=\tmu_R^2-{\tlambda_R\over 32\pi^2}\Bigl[\hat\Lambda^2-\tmu_R^2{\rm ln}\bigl(
{\hat\Lambda^2\over\tmu_R^2}\bigr)\Bigr]\qquad\qquad\qquad
\qquad\qquad~~~~ :O(4)~,}}
where $\Lambda$ and $\hat\Lambda$ are, respectively, the three- and
four-momentum ultraviolet cut-offs. Therefore, in the low-temperature
limit with $k^2 >> \mu^2_{\beta,k}$, $\ubk$ can be transformed to $\uhk$ with
the approximate scaling relation $k\to \hk/{\sqrt 2}$. The reason for
the existence of scaling can be explained by noting that
the one-loop zero temperature contribution to $\tilde\ubk$
from \uc\ can be rewritten at $k=0$ as:
\eqn\ucc{\tilde U^{(1)}_{k=0}(\Phi)={1\over 2}\int_0^{\Lambda}
{{d^3{\s p}}\over (2\pi)^3}\sqrt{{\s p}^2+V''(\Phi)}
={1\over2}\int_k^{\infty}{d^3\s p\over(2\pi)^3}\int_{-\infty}^\infty
{dp_0\over2\pi}~{\rm ln}\Bigl[p_0^2+{\s p}^2+V''(\Phi)\Bigr],}
where we have used
\eqn\mkf{\int^{\infty}_{-\infty}{dp_0\over 2\pi}
{}~{\rm ln}\bigl[ p_0^2+E^2\bigr]=E,}
which holds up to an $E$-independent constant. With the coordinate
transformation
\eqn\coor{\cases{\eqalign{p_0&=p ~{\rm sin}\theta \cr
\s p&=p ~{\rm cos}\theta \cr}}}
and imposing the same cut-off regulator $\Lambda$ for $p_0$ such that
$0 \le p^2 \le 2\Lambda^2$, the above
expression becomes:
\eqn\ucy{\eqalign{\tilde U^{(1)}_{k=0}(\Phi)&={1\over 4\pi^2}
\int_0^{\Lambda}d{\s p}
{\s p}^2\int_0^{\Lambda}dp_0~{\rm ln}\Bigl[p_0^2+{\s p}^2+V''(\Phi)\Bigr] \cr
&
={1\over 4\pi^2}\int_0^{\pi/2}d\theta~{\rm cos}^2\theta\int_0^{\sqrt{2}\Lambda}
dp~p^3{\rm ln}~\Bigl[p^2+V''(\Phi)\Bigr] \cr
&
{1\over 16\pi^2}\int_0^{\sqrt{2}\Lambda}dp~p^3~{\rm ln}\Bigl[p^2
+V''(\Phi)\Bigr] \cr
&
={1\over 2}\int_0^{\sqrt{2}\Lambda}{d^4p\over (2\pi)^4}~{\rm ln}\Bigl[p^2
+V''(\Phi)\Bigr] .}}
Therefore, we see that in the UV regime where $k\approx\Lambda$,
with the rescaling $\Lambda={\tilde\Lambda}/{\sqrt{2}}\to\infty$,
the results derived from $O(3)$ and $O(4)$ RG schemes agree with each other.

On the other hand, in the IR limit where $k^2 << \mu_k^2$ and
$\hk^2 << \hat\mu_{\hk}^2$, we have
\eqn\infrg{\cases{\eqalign{\tilde\beta_2&=-{\tlambda_Rk^3\over{16\pi^2\tmu_R}}
\Bigl(1-{1\over 2}{k^2\over{\tmu_R^2}}\Bigr)+\cdots \qquad\qquad\bigl[{\rm
O(3),~IMA}\bigr]\cr
\beta_2&= -{\lambda_kk^3\over 8\pi^2\mu_k}
\Bigl(1-{1\over 2}{k^2\over\mu_k^2}\Bigr)+\cdots \qquad\qquad~~~\bigl[{\rm
O(3),~RG}\bigr]\cr
\hat{\hat\beta}_2&=-{\hat\lambda_R\hk^4\over 16\pi^2\hat\mu^2_R}
\Bigl(1- {\hk^2\over \hat\mu^2_R}\Bigr)+\cdots \qquad\qquad~~~\bigl[{\rm
O(4),~IMA}\bigr]\cr
\hat\beta_2&=-{\hat\lambda_{\hk}\hk^4\over 16\pi^2\hat\mu^2_{\hk}}
\Bigl(1- {\hk^2\over \hat\mu^2_{\hk}}\Bigr)+\cdots \qquad\qquad~~~~\bigl[
{\rm O(4),~RG}\bigr],\cr}}}
and
\eqn\infrgo{\cases{\eqalign{\tilde\beta_4&={3\tlambda_R^2k^3\over
16\pi^2\tmu_R^3}\Bigl(1-{3k^2\over{2\tmu_R^2}}\Bigr)+\cdots
\qquad\qquad\qquad\qquad\qquad~\bigl[{\rm O(3),~IMA}\bigr]\cr
\beta_4&={3\lambda_k^2k^3\over 16\pi^2\mu_k^3}
\Bigl(1-{3k^2\over 2\mu_k^2}\Bigr)
-{g_kk^3\over{8\pi^2\mu_k}}\Bigl(1-{k^2\over 2\mu_k^2}\Bigr)+\cdots
\qquad\bigl[{\rm O(3),~RG}\bigr] \cr
\hat{\hat\beta}_4&={3{\hat\lambda}^2_R\hk^4\over 16\pi^2\hat\mu^2_R}
\Bigl(1-{2\hk^2\over \hat\mu^2_R}\Bigr)+\cdots
\qquad\qquad\qquad\qquad\qquad~~\bigl[{\rm O(3),~IMA}\bigr]\cr
\hat\beta_4&={3{\hat\lambda}^2_{\hk}\hk^4\over 16\pi^2\hat\mu^2_{\hk}}
\Bigl(1-{2\hk^2\over \hat\mu^2_{\hk}}\Bigr)
-{{\hat g_{\hk}}k^4\over 16\pi^2\hat\mu^2_{\hk}}\Bigl(1-{\hk^2\over
\hat\mu^2_{\hk}}\Bigr)+\cdots \qquad\bigl[{\rm O(4),~RG}\bigr].\cr}}}
Comparing \infrg\ and \infrgo, one sees that the running parameters
increase more rapidly for the $O(3)$ symmetric case near $k\approx 0$
by a factor of ${\mu_k/k}$ in the IR regime, and the scaling relation
between $k$ and $\hk$ for transforming $\ubk$ and $\uhk$ in the UV
regime is clearly lost. This is due to the fact that \mkf\ no longer
holds for finite $k$ and that a transformation similar to that in \ucy\
for finite $k$ ceases to exist. The numerical results for the flow equations
in \hemo\ and \hbet\ are illustrated in Figs. 5 and 6.
For the mass parameters shown in Fig. 5, the flow begins at $\Lambda
=\hat\Lambda/{\sqrt{2}}$ where all schemes take on the same negative bare mass
parameter $\mu_B^2$. We have chosen $\mu_B^2 <0$ although in principle
it can also be positive. However, the physics we are interested lies in
the IR regime which is very far from the cut-off. Therefore, the low
energy physics is influenced by the sign of the renormalized mass parameter
but not of $\mu_B^2$. The scaling relation between
the $O(3)$ and the $O(4)$ schemes in the large $k$ limit is seen by
noting that the two curves are approximately parallel to each other initially.
One also observes that both RG improved prescriptions
yield the same mass parameter at $k=0$, independent of the underlying
symmetry of the RG. Such a behavior is also observed within the
IMA schemes. This symmetry-independent behavior is again due to the
transformation \ucy\ which leads to identical one-loop effective potential.
The same argument can also be made for the RG improved prescriptions by
simply replacing $V''(\Phi)$ in \ucy\ by the corresponding
$U_{\beta,k}''(\Phi)$.
This also explains the necessity of having a more rapid convergence in
the deep IR limit in order that both RG schemes yield the same improved
parameters.

Another notable feature arises from comparing IMA with
the RG approach, where we see that by choosing an initial condition
$\mu_B^2<0$, $\mu_{k=0}^2/\mu_R^2\approx 4.6$ for $\Lambda=7$. Such an
enhancement in the mass parameter originates from the systematic
accumulation of higher loop contributions in the course of repeated blocking
transformation which brings the theory to the IR regime. On the other
hand, if $\mu_B^2>0$ is chosen, the RG improved $\mu_{k=0}^2$ will be
comparable to the IMA result $\mu_R^2$.

\ifnum\dofig=1
\bigskip

\centerline{\epsfbox{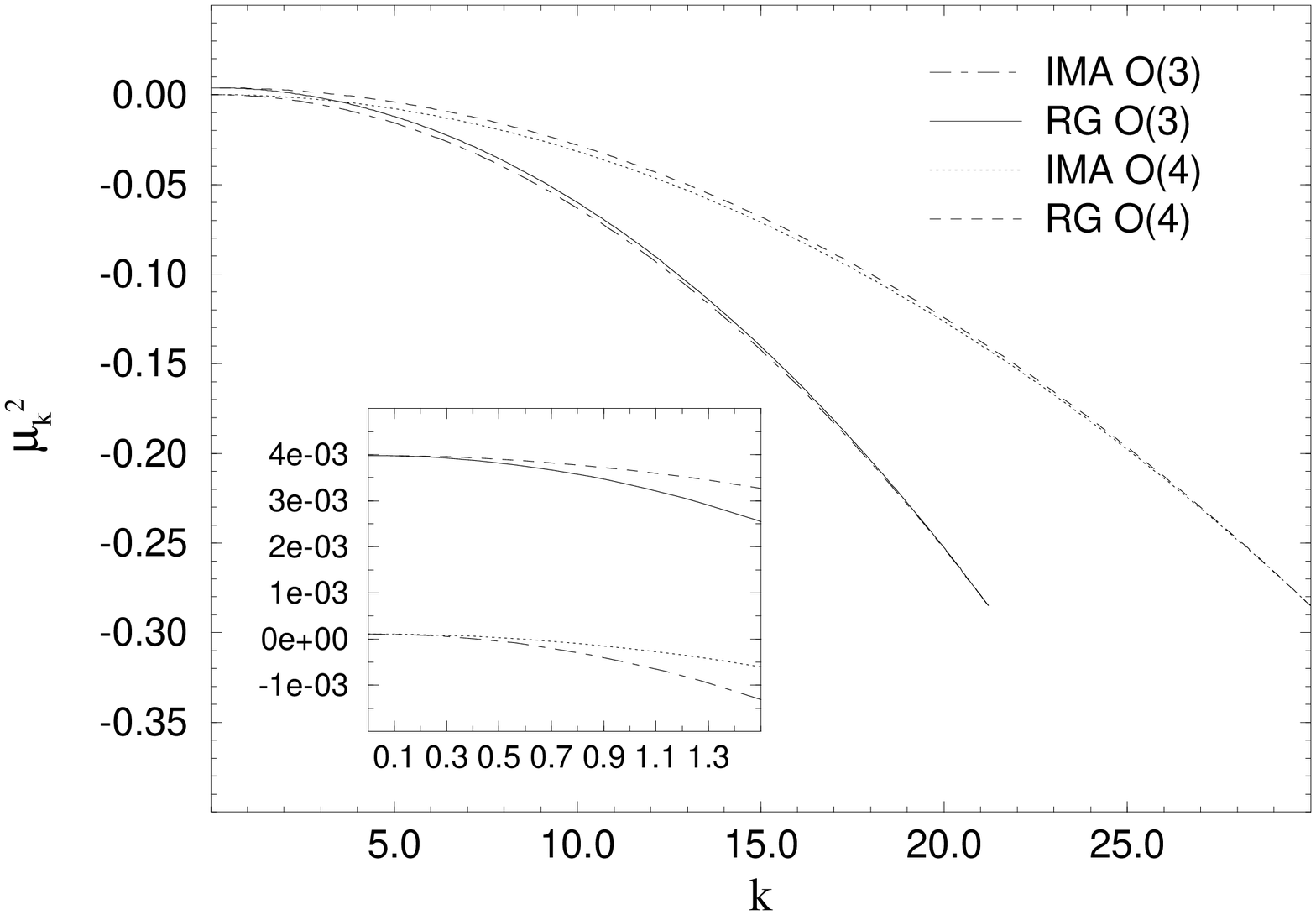}}
\bigskip
{\narrower
{\sevenrm
{\baselineskip=7pt
\itemitem{Figure 5.}
Comparison of the flow of the mass parameter
at $\scriptstyle T=0$ for different RG prescriptions using
$\scriptstyle \tilde\mu_R^2=10^{-4}$, $\scriptstyle \tilde\lambda_R=0.1$,
$\scriptstyle \tilde{\Lambda}=30$, and $\scriptstyle \Lambda =
\tilde{\Lambda}/\sqrt{2}$.
\bigskip
}}}
\fi

The remarks on the running of the mass parameter also apply qualitatively
to the flow pattern of the coupling strength which is displayed in Fig. 5.
However, contrary to the former, there exists only a
minute difference between the RG and IMA results. Therefore,
the coupling constants obtained with various approaches
near zero temperature are approximately equal. As we shall see in the next
section, this is not the case for the large $T$ limit, a regime where
the IMA scheme becomes unreliable.

\ifnum\dofig=1
\bigskip

\centerline{\epsfbox{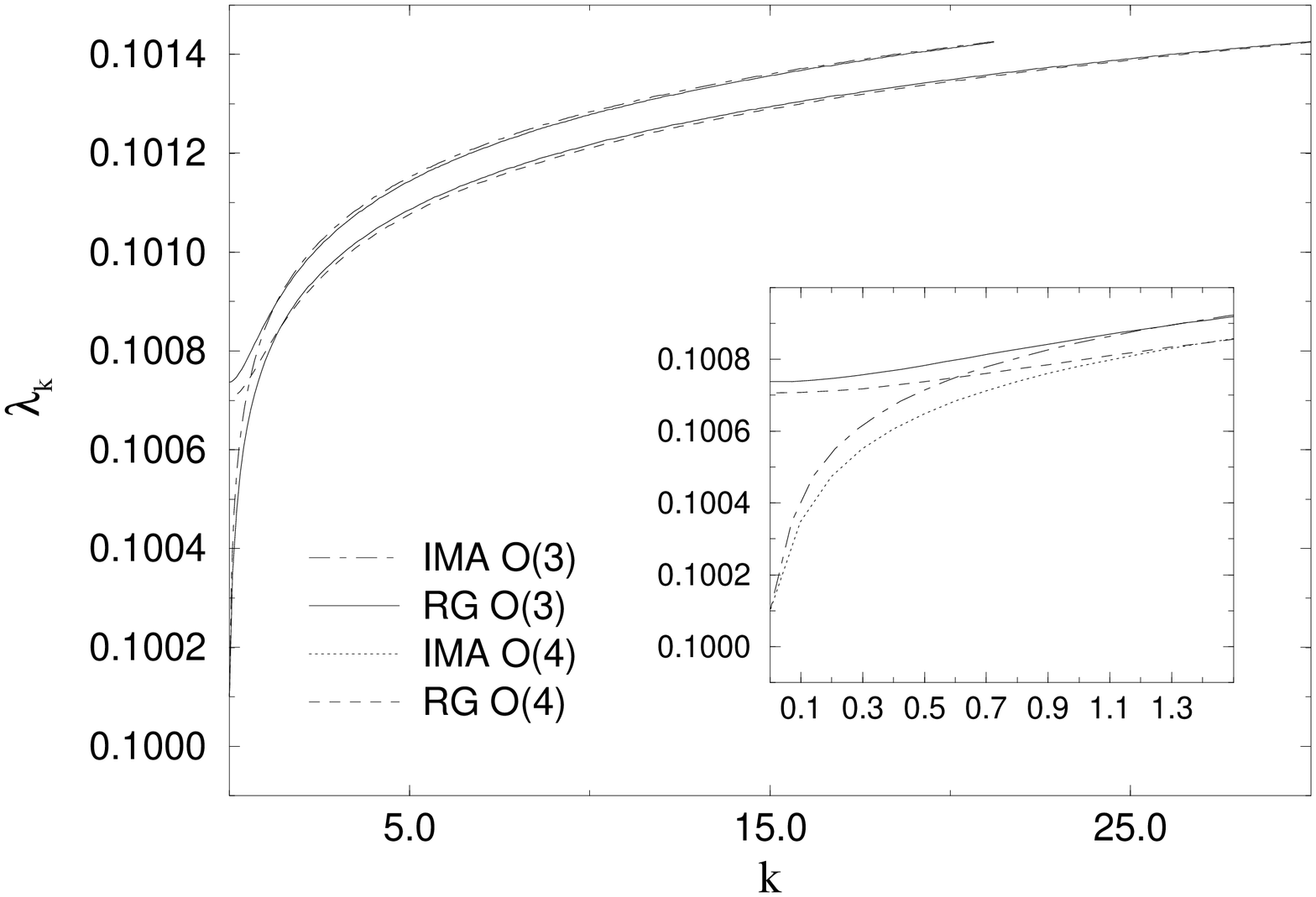}}
\bigskip
{\narrower
{\sevenrm
{\baselineskip=7pt
\itemitem{Figure 6.}
Comparison of the flow of the coupling constant
at $\scriptstyle T=0$ for different RG prescriptions using
$\scriptstyle \tilde\mu_R^2=10^{-4}$, $\scriptstyle \tilde\lambda_R=0.1$,
$\scriptstyle \tilde{\Lambda}=30$, and $\scriptstyle \Lambda =
\tilde{\Lambda}/\sqrt{2}$.
\bigskip
}}}
\fi

\medskip
\medskip
\centerline{\bf IV. HIGH TEMPERATURE LIMIT}
\medskip
\nobreak
\xdef\secsym{4.}\global\meqno = 1
\medskip
\nobreak

We consider next the high temperature limit where
$T>>\sqrt{k^2+U''_{\beta,k}}$. Since in this temperature range the second
term in \rgft\ dominates, if one neglects the first term in
\rgft\ entirely, the RG equation is simplified to
\eqn\rgfth{k{{\partial U_{\beta,k}(\Phi)}\over {\partial k}}=
-T{k^3\over4\pi^2}{\rm ln}\Bigl[{{k^2+U''_{\beta,k}(\Phi)}\over
{k^2+U''_{\beta,k}(0)}}\Bigr],}
or
\eqn\thrrg{k{\partial{\ol U_k(\ol\Phi)}\over{\partial k}}
=-{k^3\over4\pi^2}{\rm ln}
\Bigl[{{k^2+{\ddot {\ol U}}_k(\ol\Phi) }\over
{k^2+{\ddot{ \ol U}}_k(0)}}\Bigr],}
where ${\ol U}_k(\ol\Phi)=\beta U_{\beta,k}(\Phi)$ and the dots denote
differentiation with respect to $\ol\Phi=\sqrt{\beta}\Phi$, the new field
variable defined in $d=3$. One therefore
concludes that the theory in this limit has undergone a dimensional
reduction (DR) with temperature being completely decoupled, leaving an
effective three-dimensional theory described by \thrrg. This is precisely
the flow equation one would obtain for a three dimensional theory at $T=0$.
In other words, the high $T$ behavior of the theory in $d=4$ corresponds
to that of $d=3$ at zero temperature. Physically this
phenomenon can be explained by noting that the high temperature limit
is characterized by the shrinking of the
``imaginary time'' dimension having a period $\beta$. This in turn
implies a suppression of the $n\not=0$ non-static modes
in the Matsubara summation, thereby giving vanishing contribution to
$\ubk$. By neglecting the non-static modes completely, the remaining static
sector is what the three-dimensional theory parameterized
by ${\ol U}_k(\ol\Phi)$ describes. Notice that the coupling constant
would become $\bar\lambda_R=\lambda_R T$ carrying the dimension of mass.

The above analysis demonstrates that such a dimensional reduction
takes place only if $(k^2+U''_{\beta,k}(\Phi))/T^2 \to 0$, where
$U''_{\beta,k}(0)$ is the thermal mass parameter $\mu_{\beta,k}^2$.
The condition also implies a small momentum scale $k << T$.
However, as noted in \ref\landsman, DR strictly does not takes place
in the infinite
temperature limit because $\mu_{\beta,k}^2$ acquires a $T^2$-dependent
correction in the leading order which renders the ratio
$U''_{\beta,k}(\Phi)/T^2$ finite for all $T$. In Fig. 7,
we compare the parameters obtained from \thrrg\ with that of the
full RG equation \rgft. In both mass and coupling constant,
we see that the two results differ by a finite constant gap which
persists to arbitrary large value of $T$. This indeeds supports
the conclusion drawn in \landsman. Nevertheless, for sufficiently
large $T$, the theory exhibits a ``partial'' dimensional reduction
since the minute difference can be neglected. We also notice that the value of
$\lambda_{\beta}$ at $T=0$ predicted by the DR prescription
corresponds to the bare coupling constant
$\lambda_{B}>\tilde\lambda_R$. One must remember, however, that the results
obtained from the dimensioanlly reduced prescription for low $T$ are not
reliable.

\ifnum\dofig=1
\bigskip

\centerline{\epsfbox{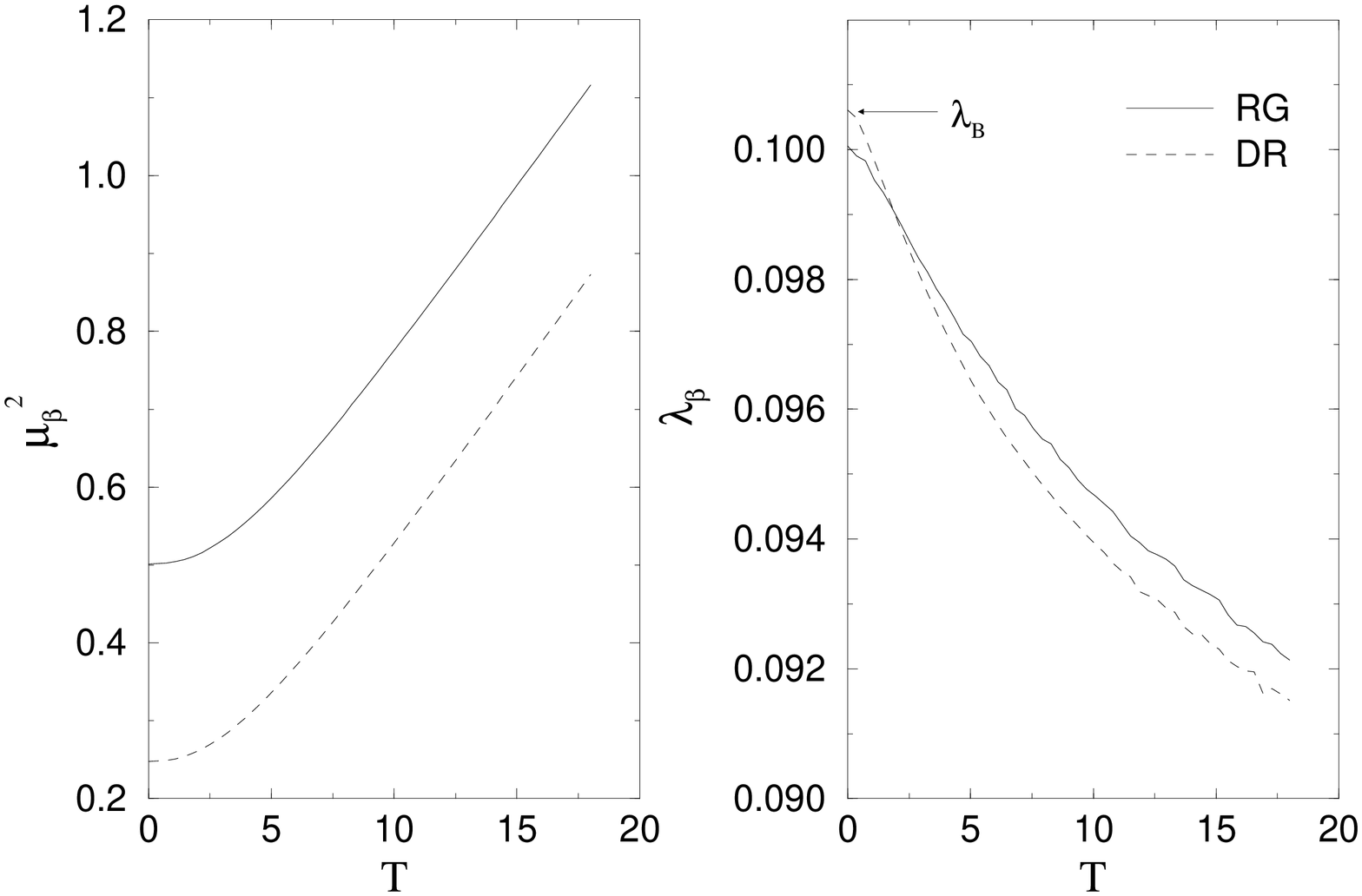}}
\bigskip
{\narrower
{\sevenrm
{\baselineskip=7pt
\itemitem{Figure 7.}
Limitation on dimensional reduction at
high temperature. Notice the gap between the mass parameters generated from the
dimensionally reduced and the full RG prescriptions.
\bigskip
}}}
\fi

Another interesting issue one may explore in the high temperature limit
is to examine the behavior
of the effective thermal coupling constant $\lambda_{\beta,k}$
\ref\babu\ and demonstrate the ineffectiveness of the one-loop IMA scheme.
To see how perturbation theory breaks down, we notice first that from
the IMA, the effective scale-dependent thermal parameters can be
written as \lpt:
\eqn\efm{\eqalign{\tilde\mu_{\beta,k}^2&=\tilde\mu_k^2
+{\tilde\lambda_R\over 4\pi^2\beta^2}\int_{\beta\sqrt{k^2+\tilde
\mu_R^2}}^{\infty}~dx{{\sqrt{x^2-\beta^2\tilde\mu_R^2}}\over {e^x-1}} \cr
&
=\tilde\mu_k^2+{\tilde\lambda_R\over 4\pi^2\beta^2}\int_0^{\infty}
dz~{z\sqrt{z^2+b^2}\over\sqrt{z^2+a^2}}\Bigl(e^{\sqrt{z^2+a^2}}-1\Bigr)^{-1},}}
and
\eqn\efc{\eqalign{\tilde\lambda_{\beta,k}&=\tilde\lambda_k
-{3\tilde\lambda_R^2\over 8\pi^2}\int_{\beta\sqrt{k^2+\tilde\mu_R^2}}^{\infty}
{}~dx{{\sqrt{x^2-\beta^2\tilde\mu_R^2}}\over x^2}~{{e^x-1+xe^x}\over
{(e^x-1)^2}} \cr
&
=\tilde\lambda_k-{3\tilde\lambda_R^2\over 8\pi^2}\int_0^{\infty}
dz~{z\over\sqrt{\bigl(z^2+a^2\bigr)\bigl(z^2+b^2\bigr)}}
\Bigl(e^{\sqrt{z^2+a^2}}-1\Bigr)^{-1},}}
where $a^2=\beta^2(k^2+\tilde\mu_R^2)$ and $b^2=\beta^2k^2$ and the
last integral expression in \efc\ is obtained via an integration
by part. In the above,
\eqn\efmm{\eqalign{\tilde\mu_k^2&=\tilde\mu_R^2-{\tilde\lambda_R\over
{64\pi^2\sqrt{k^2+\tilde\mu_R^2}}}\Biggl\{4k^3+\tilde\mu_R^2\Bigl(3k
-\sqrt{k^2+\tilde\mu_R^2}\Bigr)-{\tilde\mu_R^4\over
{k+\sqrt{k^2+\tilde\mu_R^2}}} \cr
&\qquad\qquad
-4\mu_R^2\sqrt{k^2+\mu_R^2}
{\rm ln}\Bigl({{k+\sqrt{k^2+\mu_R^2}}\over \mu_R}\Bigr)\Biggr\}}}
and
\eqn\effcou{\tilde\lambda_k=\tilde\lambda_R
-{3\tilde\lambda_R^2\over16\pi^2}\Biggl\{{{k\bigl[2k+(2k^2+\tilde\mu_R^2)
(k^2+\tilde\mu_R^2)^{-1/2}\bigr]}\over{(k+\sqrt{k^2+\tilde\mu_R^2})^2}}
-{\rm ln}\Bigl({{k+\sqrt{k^2+\tilde\mu_R^2}}\over\tilde\mu_R}\Bigr)\Biggr\}}
represent the running parameters at $T=0$.
The integrals appearing in \efm\ and \efc\ from finite
temperature contribution can be approximated
in the limits of large or small $a$ and $b$. The details are provided in the
Appendix following the classic treatment by Dolan and Jackiw \jackiw.
With $({\rm A}.15)$ and $({\rm A}.34)$ in the Appendix, the one-loop
approximation for the thermal parameters in the small $a$ and $b$ limits
take on the forms
\eqn\tmas{\eqalign{\tmk&=\tilde\mu^2_k+{{\tlambda_R T^2}\over 24}
+{\tlambda_R\tmu^2_R\over 16\pi^2}
-{{3\tlambda_R k^2}\over 16\pi^2}-{{\tlambda_R\tmu_R T}\over 4\pi^2}\Bigl[
{\pi\over 2}-{\rm sin}^{-1}\Bigl({k\over\sqk}\Bigr)\Bigr]\cr
&
-{\tlambda_R\tmu^2_R\over 16\pi^2}\Bigl[~{\rm ln}\Bigl({\sqk\over {4\pi T}}
\Bigr)+\gamma+~{\rm tanh}^{-1}\Bigl({k\over\sqk}\Bigr)\Bigr],}}
and
\eqn\tcou{\eqalign{\tlk&=\tilde\lambda_k-~{3\tlambda_R^2T\over 8\pi^2\tmu_R}
\Bigl[~{\pi\over 2}-~{\rm sin}^{-1}\Bigl({k\over\sqk}\Bigr)\Bigr]
{}~+{3\tlambda_R^2k\over 32\pi^2T} \cr
&
-{3\tlambda_R^2\over 16\pi^2}\Bigl[~{\rm ln}\Bigl({\sqk\over {4\pi T}}\Bigr)
+\gamma+~{\rm tanh}^{-1}\Bigl({k\over\sqk}\Bigr)\Bigr].}}

Eq. \tmas\ shows that at sufficiently high $T$ the mass parameter
grows quadratically with $T$. However, the presence of
${-3\tlambda_Rk^2/16\pi^2}$ tends to decrease $\tmk$. Again, we observe
a competition between the internal scale $k$ and the external parameter
$T$.

On the other hand, for large $T$, one naively expects the negative linear
$T$-dependent term to
dominate \tcou\ giving rise to a vanishing or even negative $\tlk$. However,
it is a well known fact that the correct high $T$ behavior of $\tlk$
cannot be accounted for by the simple one-loop result \tcou, and that
when higher loop contributions such as daisy and superdaisy graphs are
incorporated, the linear term will be suppressed. Since it is a
nontrivial task to resum these higher loop effects, various
attempts based on the use of gap equations \ref\gap, the Schwinger-Dyson
equation \ref\dyson\
or RG have been made to provide an effective resummation. However, the
details of resummation have raised some concerns \ref\fendley.
We shall now illustrate how
the RG equation obtained in \rgft\ can be utilized to addressed these
issues.

As explained in sec. II, a resummation of all possible non-overlapping
diagrams can be achieved with our RG equation \rgft.
By comparing \rgft\ with \rgfti, we see that the difference between the
IMA and the RG prescriptions is due to the fact that in the latter,
instead of using the $k$-independent $V(\Phi)$, $\ubk$ is employed
to give the scale-dependent thermal parameters $\tmk$ and $\tlk$
which are then used in the evaluation of $U_{\beta, k-\Delta k}
(\Phi)$. By iterating this procedure, higher loop contributions are
automatically taken into account. Guided by this logic, we proceed to
improve \tmas\ and \tcou\ by replacing the right-hand-side of the
expressions by the effective $\tmk$ and $\tlk$. This leads to the
following set of two coupled equations in the limit of small
$a=\beta(k^2+\mbk)^{1/2}$:
\eqn\gapm{\eqalign{\mbk&=\mu^2_k+{{\lbk T^2}\over 24}
+{\lbk~\mbk\over 16\pi^2}
-{{\lbk~\mk T}\over 4\pi^2}\Bigl[{\pi\over 2}-
{}~{\rm sin}^{-1}\Bigl({k\over\sqkb}\Bigr)\Bigr]\cr
&
-{{3\lbk k^2}\over 16\pi^2}
-{\lbk~\mbk\over 16\pi^2}\Bigl[~{\rm ln}\Bigl(~{\sqkb\over {4\pi T}}
{}~\Bigr)+\gamma+~{\rm tanh}^{-1}\Bigl({k\over\sqkb}~\Bigr)\Bigr],}}
and
\eqn\gapc{\eqalign{\lbk&=\lambda_k-~{3\lbk^2T\over 8\pi^2\mk}
\Bigl[~{\pi\over 2}-~{\rm sin}^{-1}\Bigl({k\over\sqkb}\Bigr)\Bigr]
{}~+{3\lbk^2k\over 32\pi^2T} \cr
&
-{3\lbk^2\over 16\pi^2}\Bigl[~{\rm ln}\Bigl({\sqkb\over {4\pi T}}\Bigr)
+\gamma+~{\rm tanh}^{-1}\Bigl({k\over\sqkb}\Bigr)\Bigr].}}
These coupled equations are actually inferred by \efmass\ and \efcou.
The only difference is that we have truncated the higher order
contributions in \gapc\ and \gapm.
Notice that they are slightly different from those
described by Chia \gap\ in that the continuous feedbacks
from $\mbk$ to $\lbk$ and vice versa are systematically incorporated,
hence leading to a more accurate determination of the
high temperature behavior of the theory.

Taking the $k=0$ limit for simplicity, we have
\eqn\gapmm{\mbb=\mu_R^2+{{\lb T^2}\over 24}
-{{\lb~\mb T}\over 8\pi}
-{\lb~\mbb\over 16\pi^2}\Bigl[~{\rm ln}\Bigl(~{\mb\over {4\pi T}}
{}~\Bigr)+\gamma-1\Bigr],}
and
\eqn\gapcc{\lb=\lambda_R-~{3\lb^2T\over 16\pi\mb}
-{3\lb^2\over 16\pi^2}\Bigl[~{\rm ln}\Bigl({\mb\over {4\pi T}}\Bigr)
+\gamma\Bigr],}
where the subscript $k$ is dropped for brevity. From \gapmm, we
notice the quadratic $T$ dependence of the thermal mass parameter $\mbb$.
To probe the high $T$ behavior of $\lb$, however, would require a more precise
determination of the ratio $\calr=\mb/T$ present in the
equation for $\lb$.
In the one-loop approximation, the use of $\mb=\mu_R$ for \gapcc\
leads to a rapid linear decrease of $\lb$ with $T$, signalling the
breakdown of perturbation theory. By including the dominant $\lb T^2/24$ term
which gives $\calr\to\lb^{1/2}/2\sqrt{6}$ as $T\to\infty$, we obtain
from \gapcc\ the relation
\eqn\rell{ \lambda_R=\lb+{3\sqrt{6}\over 8\pi}\lb^{3/2}+\cdots,}
which for $\lambda_R=0.1$ takes on the value $\lb=0.09186$.
Further refinement with $\calr={\lb^{1/2}/{2\sqrt{6}}}-{3\lb/{48\pi}}$
yields
\eqn\relr{ \lambda_R=\lb+{3\lb^2\over 16\pi}\Bigl({\lb^{1/2}\over{2\sqrt{6}}}
-{3\lb\over{48\pi}}\Bigr)^{-1/2}+\cdots,}
or $\lb=0.09160$. This is in accord with that illustrated in Fig. 8
for the $T$ dependence of the thermal parameters $\mb$ and $\lb$.
Thus, we conclude that at very high $T$, $\lb$ approaches a constant
nonvanishing
positive value. In Fig. 7, the infinitesimal decrease of $\lb$ in the
high $T$ regime is due to the nature of the coupled equations in which
a particular value of $\lb$ is first used in \gapmm\ for deducing $\calr$ which
is subsequently inserted to \gapcc\ for the determination of a new improved
$\lb$. In general, the more accurately $\calr$ is, the better
the agreement with that generated by \rgft.
We emphasize here that contrary to the claim in \fendley, $\lb$ can
never increase with $T$ because of the negative sign associated with the
temperature-dependent term in the full RG flow equation \rgft. The
inclusion of higher loop effects can only modify the leading $T$-dependent
behavior, but not the sign. When RG is invoked to study $\lambda_{\beta,k}$,
it is crucial to keep in mind that there are three dimensionless combinations
$b={k/T}$, $b'=\tilde\mu_R/k$ and $b''=\tilde\mu_R/T$ present in the flow
equation with $b'$ and $b''$ having opposite effect
to the running of $\lambda_{\beta,k}$. Erroneous use of $b''$
in the RG analysis can lead to the wrong claim that $\lambda_{\beta,k}$
rises with $T$. By employing the correct choice $b'$,
one will reproduces the standard result that $\lambda_{\beta,k}$
increases logarithmically with $k$.

\ifnum\dofig=1
\bigskip

\centerline{\epsfbox{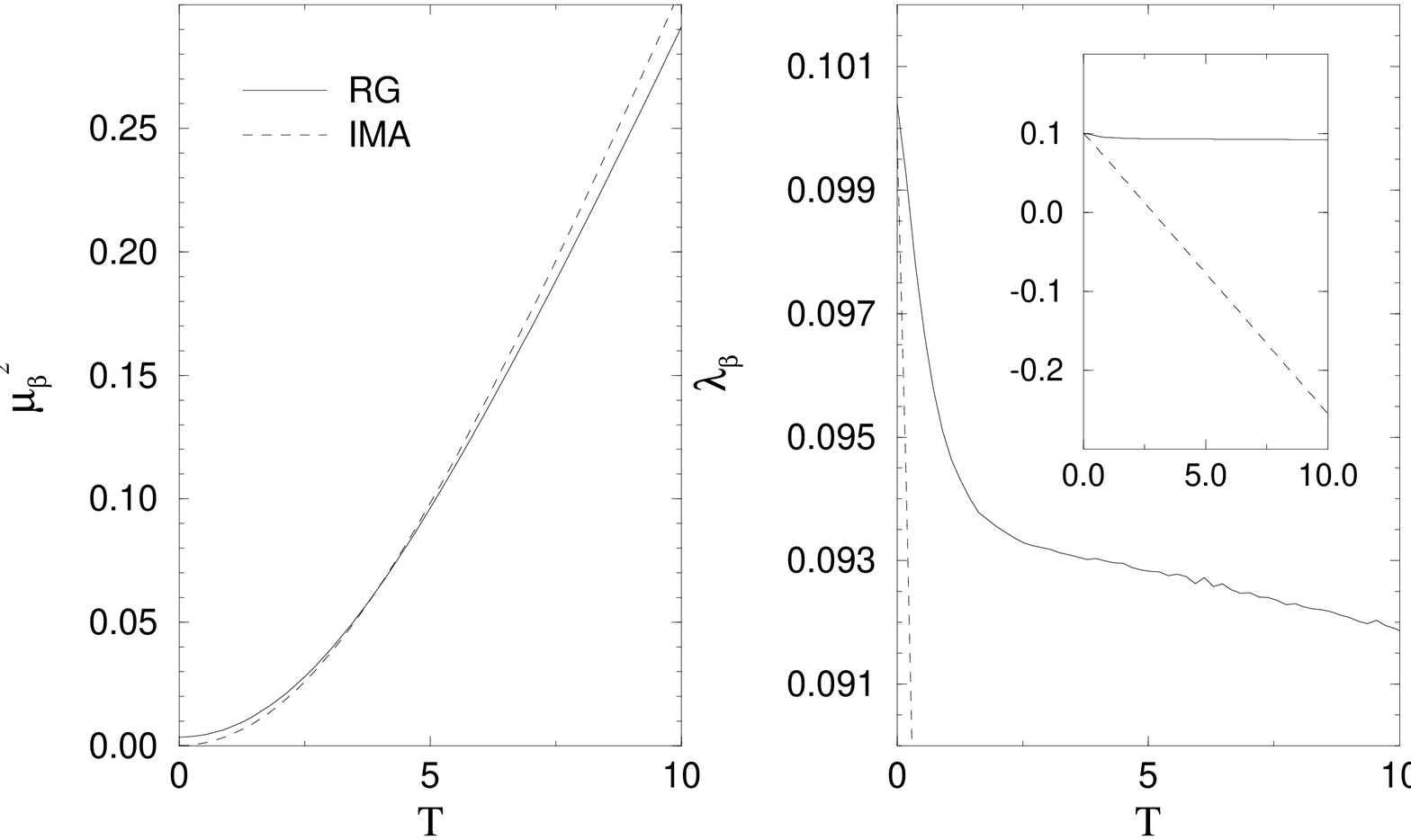}}
\bigskip
{\sevenrm
\centerline{
Figure 8.
Temperature dependence for $\scriptstyle \mu_{\beta}^2$ and $\scriptstyle
\lambda_{\beta}$ at $\scriptstyle k=0$.}
\bigskip
}
\fi

We also notice that the mass parameter $\mbb$ obtained with the RG method
is slightly smaller than that of the one-loop result. The reason again is
due to the continuous feedback of the higher loop effects as well as the
use of improved value for the thermal coupling $\lb$ which is smaller
compared to $\lambda_R$.

Since the scales $k$ and $T$ enter in a complementary manner,
in the regime where $k >> T$ one expects the
thermal effects to be suppressed. This can be seen by noting that
\eqn\supm{\tilde\mu_{\beta,k}^2=\tilde\mu_k^2+{\tilde\lambda_R\over
4\pi^2\beta^2}~e^{-a}\Bigl[1+{a\over 2}+{b^2\over 2a}+\cdots\Bigr],}
and
\eqn\supl{\tilde\lambda_{\beta,k}=\tilde\lambda_k-{3\tilde\lambda_R^2\over
8\pi^2}~{e^{-a}\over a}\Bigl[1+a-{1\over 2}(a^2-b^2)+\cdots\Bigr]}
with the help of $({\rm A.}19)$ and $({\rm A.}37)$. Indeed, the
thermal parameters are suppressed exponentially by a factor $e^{-a}$.
When the blocking scale $k$ coincides with the UV cut-off $\Lambda$,
all thermal contributions must vanish since for $k=\Lambda$ the original
temperature-independent bare theory is recovered. Equivalently, one may
say that the counterterms are independent of temperature. This implies
that we must have $e^{-\beta\Lambda}\to 0$ in principle.
Correspondingly, we work within a temperature range which lies
sufficiently far from $\Lambda$ to ensure that nearly all
particles will have a momentum below $\Lambda$ according to
the Bose-Einstein distribution. The suppression of thermal effects
for large $\beta k$ can indeed be seen from Fig. 4.
\medskip
\medskip
\centerline{\bf V. SPONTANEOUS BREAKING AND RESTORATION OF SYMMETRY}
\medskip
\nobreak
\xdef\secsym{5.}\global\meqno = 1
\medskip
\nobreak

Turning to the scenario of spontaneous symmetry breaking (SSB) with
$\mu_R^2 <0$, the one-loop finite temperature blocked potential becomes
\eqn\upb{\eqalign{\tilde U_{\beta,k}(\Phi)&={\tilde\mu_R^2\over2}\Phi^2
\Bigl(1-{\tilde\lambda_R\over64\pi^2}\Bigr)
+{\tilde\lambda_R\over4!}\Phi^4\Biggl[1-{\tilde\lambda_R\over 256\pi^2}
{{36\tilde\mu_R^4+84\tilde\lambda_R\tilde\mu_R^2M^2+25\tilde\lambda_R^2
M^4}\over{\bigl(\tilde\mu_R^2+\tilde\lambda_RM^2/2\bigr)^2}}\Biggr] \cr
&
+{1\over64\pi^2}\Biggl\{-2k\Bigl(2k^2+\tilde\mu_R^2+{1\over2}\tilde\lambda_R
\Phi^2\Bigr)\Bigl({k^2+\tilde\mu_R^2+{1\over2}\tilde\lambda_R\Phi^2}
\Bigr)^{1/2}\cr
&
+\Bigl(\tilde\mu_R^2+{1\over2}\tilde\lambda_R\Phi^2\Bigr)^2{\rm ln}
\Bigl[{{2k^2+\tilde\mu_R^2+\tilde\lambda_R\Phi^2/2+2k\sqrt{k^2+\tilde\mu_R^2
+\tilde\lambda_R\Phi^2/2}}
\over{\tilde\mu_R^2+\tilde\lambda_RM^2/2}}\Bigr]\Biggr\}\cr
&+{1\over2\pi^2\beta}\int_k^{\Lambda}dpp^2{\rm ln}
\Bigl[1-e^{-\beta\sqrt{p^2+\tmu_R^2+\tlambda_R\Phi^2/2}}\Bigr],}}
where we have chosen the following off-shell renormalization conditions
used in \coleman:
\eqn\hee{\cases{\eqalign{\tilde\mu^2_R&={\rm Re}~\Biggl[{\partial^2{\tilde
U_{\beta,k}}\over\partial\Phi^2}\Big\vert_{\Phi=\beta^{-1}=k=0}\Biggr]~ < 0\cr
\tilde\lambda_R&={\partial^4{\tilde U_{\beta,k}}\over\partial\Phi^4}
\Big\vert_{\Phi=M,\beta^{-1}=k=0,} ~,\cr}}}
where $M=<\Phi>$ is the nonvanishing vacuum expectation at which
$\tilde U_{\beta,k}(\Phi)$ is minimized.
Since $\tilde\mu_R^2 < 0$, it can no longer be
interpreted as the mass parameter for the theory. Instead, the mass
parameter is determined by
$\tilde U''_{\beta^{-1}=k=0}(\Phi=M)\approx -2\tilde\mu_R^2 >0$.
A characteristic feature of SSB is the
development of an imaginary sector in $\tilde U_{\beta,k}(\Phi)$
in the regime where $k$ and $\Phi$ are small and the argument
$k^2+\tilde\mu_R^2+\tilde\lambda_R\Phi^2/2$
inside the logarithm and the square root becomes negative.
Any attempt to extract from the low temperature range
the critical temperature $T_c$ beyond which
symmetry restoration takes place using \upb\ is fruitless since the
resulting $T_c$ is complex due to the presence of an imaginary part in the
finite temperature contribution found in the last line of \upb\ \jackiw.
However, concentrating only on the real part of $\ubk$ and taking
into account the higher order daisy and super-daisy graphs, Dolan and
Jackiw \jackiw\ obtained a critical temperature $\tilde T_c$
\eqn\tc{\tilde T_c=\Bigl({-24\tilde\mu_R^2\over\tilde\lambda_R}\Bigr)^{1/2}}
beyond which symmetry restoration takes place via a second order phase
transition. Nevertheless, a {\it true} symmetry restoration
should be accompanied by the disappearance and not
negligence of the imaginary contribution in the thermal blocked potential
$\ubk$. This is clearly not possible within the context of the IMA
since the finite temperature effects never enter the argument
$k^2+\tilde\mu_R^2+\tilde\lambda_R\Phi^2/2$, and we see that the
imaginary contribution persists for all $T$ unless higher loop effects
daisy and super-daisy are properly taken into account \jackiw.

It is interesting to compare our finite
temperature RG prescription with that used in \jackiw\ since in our
approach the argument inside the
logarithm and square root becomes $k^2+U''_{\beta,k}(\Phi)$, and is
a positive quantity beyond $T_c$.

\ifnum\dofig=1
\bigskip

\centerline{\epsfbox{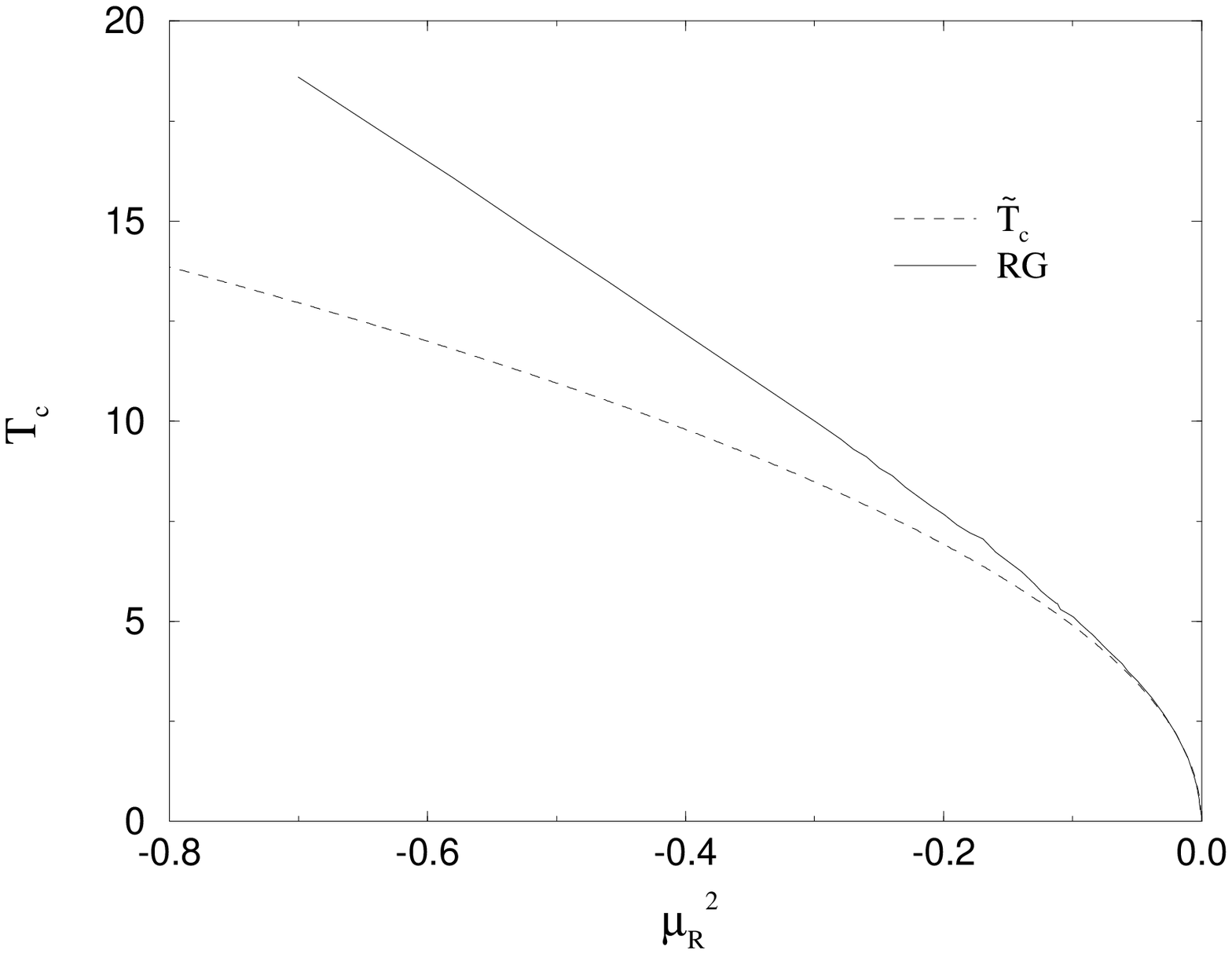}}
\bigskip
{\narrower
{\sevenrm
{\baselineskip=7pt
\itemitem{Figure 9.}
Comparison of the critical temperature $T_c$ obtained
by RG with $\tilde T_c$. Notice that a higher value is predicted for RG.
\bigskip
}}}
\fi

With the RG flow equation \rgft, we define $T_c$ to be the symmetry
restoration temperature above which the imaginary contribution vanishes
entirely. A comparison between $T_c$ and $\tilde T_c$ is
illustrated in Fig. 9. Notice that in the RG scheme, $\mu^2_R$ is used
instead of $\tilde\mu_R^2$ which together with $\tilde\lambda_R$ determines
$\tilde T_c$. For sufficiently low $-\mu_R^2$,
one finds agreement between the RG result and \tc.
However, they begin to deviate for large $-\mu_R^2$ and the
RG approach gives a higher symmetry restoration temperature compared
to $\tilde T_c$ in this limit. This observation can be explained by
noting that for a given $T$, the RG prescription yields a lower value
for the thermal mass parameter $\mu_{\beta}^2$ compared to that of \jackiw\
for large $T$ due to the fact that we also take into consideration
the evolution of the coupling constant $\lambda_{\beta}$.
Therefore, for the condition $\mu^2_{\beta_c}=0$ to be satisified,
we must have $T_c \ge \tilde T_c$. Had we replaced $\lambda_{\beta}$
by $\lambda_R$ by neglecting the influence of the running of the coupling
constant on the mass gap equation \gapmm, we would have $T_c=\tilde T_c$
instead.

Notice that it is also possible to
extract $T_c$ without confronting the complication of imaginary contribution
by starting from a temperature value $T > T_c$ and gradually
lower the temperature. Both methods lead to the same $T_c$, as has been
checked numerically. Near the critical point $T\gsim T_c$, one obtains from
\gapmm:
\eqn\muer{ \mb \approx {{\pi T}\over 3}\Bigl[1-
{1\over {\lambda_{\beta}T^2}}\lim_{T\to T_c}(\lambda_{\beta}T^2)\Bigr],}
such that $\mu_{\beta_c}=0$ at exactly $T=T_c$.
Similarly, a resummation of higher physical loop contributions in \gapcc\
shows that in the vicinity of critical point, the coupling constant
behaves as:
\eqn\laee{\lb \approx {\lambda_R\over{1+{3\lambda_R\over 16\pi}{T\over\mb}}}
\longrightarrow {16\pi^2\over 9}\Bigl[1-{1\over {\lambda_{\beta}T^2}}
\lim_{T\to T_c}(\lambda_{\beta}T^2)\Bigr],}
after substituting \muer. One therefore sees that the effective thermal
coupling constant $\lambda_{\beta}$
vanishes at the transition temperature and the theory becomes
non-interacting. This can also be obtained directly from \efc\ which
implies
\eqn\lrt{ {3{\cal I}\over 8\pi^2}\lambda_{\beta_c}^2+\lambda_{\beta_c}
-\lambda_R=0}
at precisely $T=T_c$, where
\eqn\inng{ {\cal I}=\int_0^{\infty}{dz\over{z\bigl(e^z-1\bigr)}}.}
That ${\cal I}$ diverges naturally gives $\lambda_{\beta_c}=0$.
The temperature dependences of $<\Phi>$, $\mu_{\beta}^2$ and $\lambda_{\beta}$
above and below $T_c$ are depicted in Fig. 10. Our numerical result
indeed yields $\lambda_{\beta_c}=0$ when approaching $T_c$ both from above
and below. However, for $T\lsim T_c$ in the SSB phase, there exists certain
numerical fluctuations which nevertheless diminish at $T_c$.
A second order phase
transition characterized by a continuous decrease of
the vacuum condensate is predicted with our RG flow equation.
With $\lambda_{\beta_c}=0$ at the transition
point, the IR divergence is completely lifted. Had we not incorporated
the higher loop effects, the IR singularity would persist and result in
a breakdown of the perturbation theory.

The inapplicability of the IMA scheme near $T_c$ can be illustrated by
studying the order of transition it predicts. For
$T\gsim T_c$, with $\mu_{\beta}^2\gsim 0$ and $\lambda_{\beta}<0$,
a first order transition is obtained due to a positive higher order
$\Phi^6$ contribution, in contradiction with that predicted with RG.
Our analyses are in agreements with that obtained in
\wetterich\ and \gap. However, in \wetterich\ where a smooth momentum
regulator is used, there exists residual dependence of the
running parameters on the shape of the momentum regularizing function.

Physically the critical temperature corresponds to a fixed point
in the RG trajectory. Since this fixed point is of Gaussian nature with
$\mu_{\beta_c}^2=\lambda_{\beta_c}=0$, any interaction between the
scalar fields must be of higher order and parameterized by the irrelevant
operators classified around the fixed point.
For this theory, the critical exponents can be accurately determined and shown
to coincide with that of the three-dimensional theory at $T=0$ \wetterich.
{}From Fig. 10, we also observe that after symmetry restoration with
$T > T_c$, $\lambda_{\beta}$ rises again
and eventually approaches the same constant value as that without
going through phase transition.

\ifnum\dofig=1
\bigskip

\centerline{\epsfbox{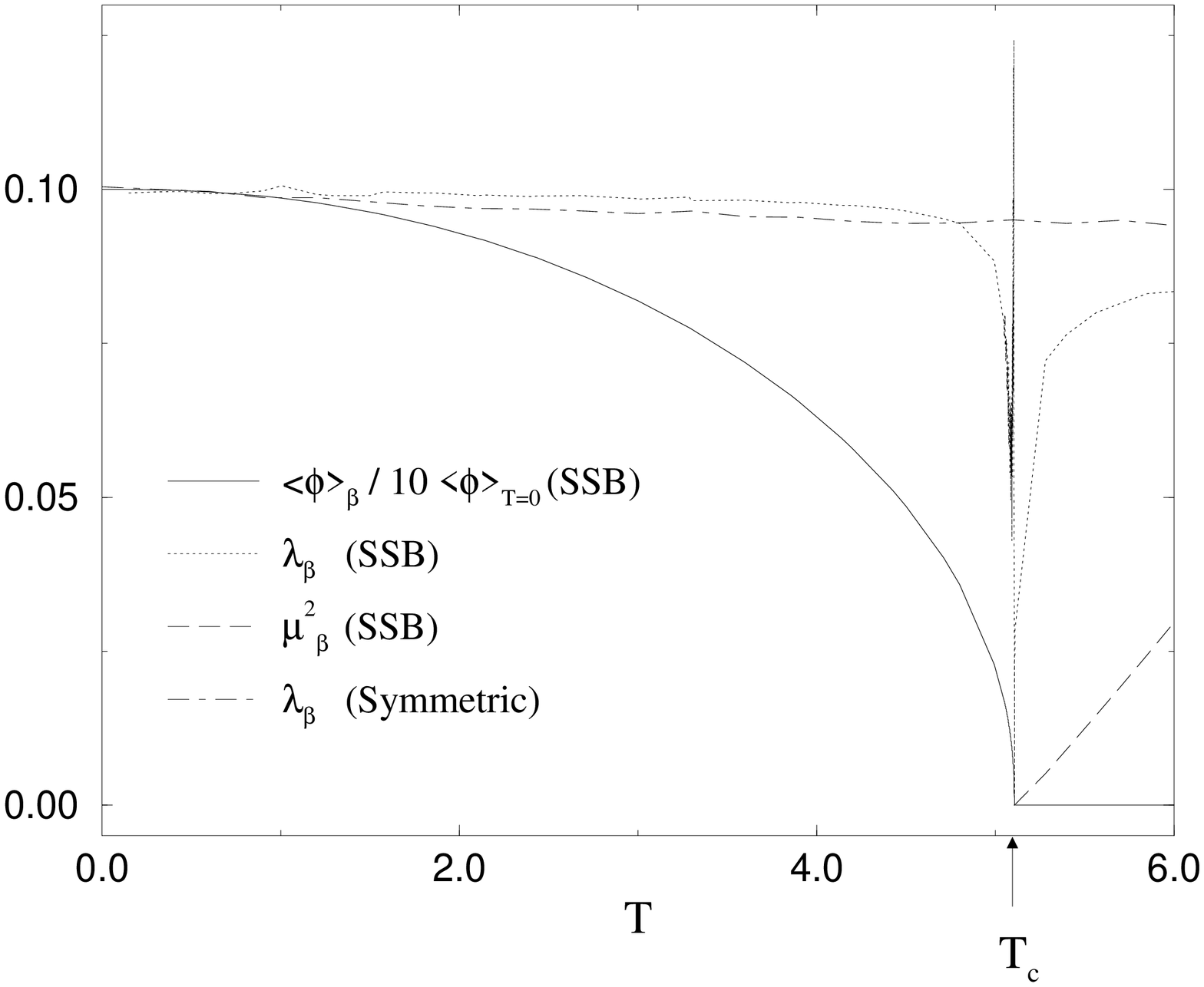}}
\bigskip
{\narrower
{\sevenrm
{\baselineskip=8pt
\itemitem{Figure 10.}
Temperature dependence of runnig parameters in the presence of spontaneous
symmetry breaking.
\bigskip
}}}
\fi

\medskip
\medskip
\centerline{\bf VI. SUMMARY AND DISCUSSIONS}
\medskip
\nobreak
\xdef\secsym{6.}\global\meqno = 1
\medskip
\nobreak

In this paper we have carried out the investigation of finite
temperature scalar theory using an improved RG program. Our RG flow
equation has successfully reproduced the characteristic behaviors
of the system in both high and low $T$ limits. In particular,
we have established a connection between the $O(4)$ and $O(3)$ blocked
potentials in large $k$ and small $k$
regimes in sec III, and in sec IV dimensional reduction and the
proper high temperature RG flow patterns for $\mu_{\beta,k}^2$ and
$\lambda_{\beta,k}$ are deduced. At $k=0$ the coupling constant
$\lambda_{\beta,k}$ is seen to decrease with $T$ and approximately
approaches a positive constant value.
Thermal effects are suppressed
for large $k$ and vanish entirely at $k=\Lambda$ where the bare
theory is defined. In addition we demonstrated in Sec. V
how true symmetry restoration is attained with vanishing imaginary
contribution in $\ubk$.

Our method is more advantageous compared to the approaches mentioned
in the Introduction in a number of ways: The coupled equations \gapm\ and
\gapc\ derived from \rgft\ for $\mu_{\beta,k}^2$ and $\lambda_{\beta,k}$
have provided  further improvement compared to that derived
for $\mu^2_{\beta}$ alone
in \jackiw\ and \gap\ since the continuous feedbacks between
$\mu_{\beta,k}^2$ and $\lambda_{\beta,k}$ are retained systematically.
In addition, through \rgft, we are able to analyze simultaneously the
competing effects of two scales, namely, $k$ and $T$ in an unambiguous manner.
A RG approach based on the running of an internal scale $k$ is more
``physical'' than that employed in \elmfors\ with the external $T$ as
the running parameter.
When applied to the Yang-Mills theory with a momentum-dependent
gluon polarization tensor, the resummation using the latter scheme is
inconsistent \elmfors.
This poses no difficulty with RG formulation which naturally includes
both the temperature and momentum effects.

Various interesting issues can be explored with our RG scheme
in light of its success. For example, one can can use this approach
to study the nature of the phase transition for the electroweak theory.
For $\lambda_R << g_R^2$, where $g_R$ is the coupling constant
for the gauge fields, one would expect a first order transition which is
required for explaining the asymmetry of the
baryogenesis \ref\kuzmin. For the Yang-Mills theory, a RG flow equation
similar to \rgft\ will provide information on the roles of $k$ and $T$
on the running of the gauge coupling constant. The flow of the theory with
at $T=0$ has been worked out \ref\gauge. It would be interesting
to investigate the effect of $T$ on such a theory which is known to
exhibit asymptotic
freedom at $T=0$. If $T$ and $k$ can generate opposite effects as for
the scalar theory, there will be nontrivial consequences on the picture
of deconfinement transition of quarks and gluons. In addition, the
resummation of ``hot thermal'' loops using this RG approach will readily
yield the gauge-independent gluon damping rate and be compared
with that obtained in \bp\ via an effective action.
Works along these directions are currently in progress.

\medskip
\medskip
\goodbreak
\bigskip
\centerline{\bf ACKNOWLEDGEMENTS}
\medskip
\nobreak
We are grateful to L. Dolan, B. M\"uller and J. Polonyi for stimulating
discussions and helpful comments on the manuscript.  This work was
supported in part by the U. S. Department of Energy (Grant No.
DE-FG05-90ER40592)
and by a grant from the North Carolina Supercomputing
Center.

\medskip
\medskip
\centerline{\bf APPENDIX}
\medskip
\nobreak
\xdef\secsym{{\rm A}.}\global\meqno = 1
\medskip
\nobreak
We proceed to evaluate the integrals shown in \efm\ and \efc\ following
the techniques utilized in \jackiw. The presence of the blocking scale
$k$ will slightly complicate the algebra compared with the previous
studies. We first consider the
small $a$ limit of the integral in \efc\ having the form
\eqn\evi{\eqalign{ I(a,b)&=\int_0^{\infty}
dz~{z\over\sqrt{\bigl(z^2+a^2\bigr)\bigl(z^2+b^2\bigr)}}
\Bigl(e^{\sqrt{z^2+a^2}}-1\Bigr)^{-1} \cr
&
={1\over 2}\int_0^{\infty}dz~{z\over\sqrt{\bigl(z^2+a^2\bigr)\bigl(z^2
+b^2\bigr)}}\Biggl[{\rm coth}~\Bigl({\sqrt{z^2+a^2}\over 2}\Bigr)-1\Biggr] \cr
&
={1\over 2}\int_0^{\infty}dz~{z\over\sqrt{\bigl(z^2+a^2\bigr)\bigl(z^2
+b^2\bigr)}}\Biggl[2\sqrt{z^2+a^2}\sum_{n=-\infty}^{\infty}
{1\over{z^2+a^2+4\pi^2n^2}}-1\Biggr] \cr
&
=I^{(1)}(a,b)+I^{(2)}(a,b)~,}}
where
\eqn\evvi{ I^{(1)}(a,b)=\int_0^{\infty}dz~{z\over\sqrt{\bigl(z^2+b^2\bigr)}}
\sum_{n=-\infty}^{\infty}{1\over{z^2+a^2+4\pi^2n^2}},}
and
\eqn\evii{ I^{(2)}(a,b)=-{1\over 2}\int_0^{\infty}dz~{z\over\sqrt{\bigl(z^2
+a^2\bigr)\bigl(z^2+b^2\bigr)}}~.}
Since divergences are encountered when splitting the integral in such a
manner, we introduce a suppression factor $z^{-\epsilon}$ to regularize
the individual sum and expect the infinities to cancel, thereby making
the final result for \evi\ finite. By rewriting \evvi\ as
\eqn\eevi{ I^{(1)}_{\epsilon}(a,b)=\sum_{n=-\infty}^{\infty}\Bigl(a^2
+4\pi^2n^2\Bigr)^{-\epsilon/2}\int_0^{\infty}dx~{x^{\epsilon}\over
{(1+x^2)\sqrt{b^2x^2+a^2+4\pi^2n^2}}},}
via a change of variable $x=(a^2+4\pi^2n^2)^{1/2}z^{-1}$, the integration
can be carried out with the help of
\eqn\expre{\eqalign{ \int_0^{\infty}dx~{x^{\epsilon}\over
{(1+x^2)\sqrt{b^2x^2+c^2}}}&={b^{1-\epsilon}c^{-2+\epsilon}\over{2\sqrt{\pi}}}
\Gamma\bigl(1-{\epsilon\over 2}\bigr)\Gamma\bigl({{-1+\epsilon}\over 2}\bigr)
F\bigl(1,1-{\eps\over 2},{3-\eps\over 2},{b^2\over c^2}\bigr) \cr
&
+{\pi~{\rm sec}\bigl(\eps\pi/2\bigr)\over 2\sqrt{c^2-b^2}},}}
where
\eqn\funn{ F\bigl(a,b,c;\gamma\bigr)=F\bigl(b,a,c;\gamma\bigr)=
B^{-1}(b,c-b)\int_0^1 dx~x^{b-1}(1-x)^{c-b-1}(1-\gamma x)^{-a},}
with
\eqn\ber{ B(x,y)={\Gamma(x)\Gamma(y)\over\Gamma(x+y)}=\int_0^1 dt~t^{x-1}
(1-t)^{y-1}.}
Substitution of \expre\ into \eevi\ then gives
\eqn\eevi{\eqalign{ I^{(1)}_{\epsilon}(a,b)&={\pi\over2}~{\rm sec}\bigl(
{\eps\pi\over2}\bigr)~\Biggl\{{a^{-\eps}\over\sqrt{a^2-b^2}}
+2\sum_{n=1}^{\infty}{{\bigl(a^2+4\pi^2n^2\bigr)^{-\eps/2}}\over\sqrt
{a^2+4\pi^2n^2-b^2}}\Biggr\} \cr
&
+{b^{1-\eps}\over{2\sqrt{\pi}}}\Gamma\bigl(1
-{\epsilon\over 2}\bigr)\Gamma\bigl({{-1+\epsilon}\over 2}\bigr)
\Biggl\{a^{-2}F\bigl(1,1-{\eps\over 2},{3-\eps\over 2},{b^2\over a^2}\bigr)\cr
&
+2\sum_{n=1}^{\infty}\bigl(a^2+4\pi^2n^2\bigr)^{-1}
F\bigl(1,1-{\eps\over 2},{3-\eps\over 2},{b^2\over{a^2+4\pi^2n^2}}\bigr)
\Biggr\},}}
which in the limit of vanishing $\eps$ becomes
\eqn\eevii{\eqalign{ I^{(1)}_{\epsilon}(a,b)&={\pi\over2}{1\over\sqrt{
a^2-b^2}}+2^{-1-\eps}\pi^{-\eps}\zeta(1+\epsilon)+{1\over2}\sum_{n=1}^{\infty}
{1\over n}\Biggl[\Bigl({1+{a^2-b^2\over{4\pi^2n^2}}}\Bigr)^{-1/2}-1\Biggr]\cr
&
-{1\over\sqrt{a^2-b^2}}~{\rm sin}^{-1}~\Bigl({b\over a}\Bigr)
-2\sum_{n=1}^{\infty}\Bigl(a^2+4\pi^2n^2-b^2\Bigr)^{-1/2}{\rm sin}^{-1}
\Bigl({b\over\sqrt{a^2+4\pi^2n^2}}\Bigr) \cr
&
={1\over 2\eps}+{\pi\over2}{1\over\sqrt{a^2-b^2}}+{1\over 2}\bigl(\gamma
-{\rm ln}~{2\pi}\bigr)-{1\over\sqrt{a^2-b^2}}~{\rm sin}^{-1}~\Bigl({b\over a}
\Bigr)-{b\over 12}+O(a^2),}}
where we have used
\eqn\zz{ \zeta(1+\eps)=\sum_{n=1}^{\infty}{1\over n^{1+\eps}}=
-{{2^{\eps}\pi^{1+\eps}\zeta(-\eps)}\over{\Gamma(1+\eps)~{\rm sin}
({\eps\pi\over2})}}=(2\pi)^{\epsilon}\Bigl[{1\over\eps}-{\rm ln}~2\pi
+\gamma+O(\eps)\Bigr],}
\eqn\hype{ F\bigl(1,1,{3\over 2},x^2\bigr)={{{\rm sin}^{-1}~x}\over
{x\sqrt{1-x^2}}},}
\eqn\ser{\eqalign{ &\sum_{n=1}^{\infty}\Bigl(a^2+4\pi^2n^2-b^2\Bigr)^{-1/2}
{\rm sin}^{-1}\Bigl({b\over\sqrt{a^2+4\pi^2n^2}}\Bigr) \cr
&
\approx b\sum_{n=1}^{\infty}\Bigl(a^2+4\pi^2n^2\Bigr)^{-1}\Bigl[1-{b^2\over
{a^2+4\pi^2n^2}}\Bigr]^{-1/2}
=b\sum_{n=1}^{\infty}\Bigl(a^2+4\pi^2n^2\Bigr)^{-1}+\cdots \cr
&
=b\Bigl[{1\over 4a}~{\rm coth}~\bigl({a\over 2}\bigr)-{1\over 2a^2}\Bigr]
={b\over 24}+{a^2b\over 1440}+\cdots,}}
and neglected
\eqn\oos{\sum_{n=1}^{\infty}
{1\over n}\Biggl[\Bigl({1+{a^2-b^2\over{4\pi^2n^2}}}\Bigr)^{-1/2}-1\Biggr]
=-{\bigl(a^2-b^2\bigr)\over 8\pi^2}\sum_{n=1}^{\infty}{1\over n^3}+
\cdots=O(a^2).}
In a similar manner, we have
\eqn\eveii{\eqalign{ I^{(2)}_{\eps}(a,b)&=-{1\over 2}\int_0^{\infty}dz~
{z^{1-\eps}\over\sqrt{\bigl(z^2+a^2\bigr)\bigl(z^2+b^2\bigr)}} \cr
&
=-{1\over 4\sqrt{\pi}}\Biggl\{a^{-1}b^{1-\eps}~\Gamma(1-{\eps\over2})
\Gamma({{-1+\eps}\over 2})~F\bigl({1\over 2},1-{\eps\over 2},{{3-\eps}\over 2},
{b^2\over a^2}\bigr) \cr
&\qquad\quad
+a^{-\eps}~\Gamma({{1-\eps}\over2})
\Gamma({\eps\over 2})~F\bigl({1\over 2},{\eps\over 2},{{1+\eps}\over 2},
{b^2\over a^2}\bigr)\Biggr\} \cr
&
\mapright{\eps\to 0}~~~ ~{1\over 2}~{\rm tanh}^{-1}\bigl({b\over a}\bigr)
-{1\over 2\eps}+{1\over 2}~{\rm ln}{a\over 2}+O(\epsilon).}}
Combining \eevii\ and \eveii\ yields the finite result:
\eqn\evy{\eqalign{I(a,b)&=\int_0^{\infty}
dz~{z\over\sqrt{\bigl(z^2+a^2\bigr)\bigl(z^2+b^2\bigr)}}
\Bigl(e^{\sqrt{z^2+a^2}}-1\Bigr)^{-1} \cr
&
={1\over\sqrt{a^2-b^2}}\Bigl[{\pi\over 2}-{\rm sin}^{-1}
{}~\Bigl({b\over a}\Bigr)\Bigr]+{1\over 2}\Bigl[{\rm ln}~{a\over 4\pi}+\gamma
+{\rm tanh}^{-1}\bigl({b\over a}\bigr)\Bigr]-{b\over 12}+O(a^2)~,}}
which for vanishing $b$, reduces to that obtained in \jackiw.

On the other hand, for $k >> T >> \mu_R$, the original integral expression
in \efc\ can be approximated as:
\eqn\sro{\eqalign{\tilde I(a,b)&=\int_a^{\infty}~dx{{\sqrt{x^2-(a^2-b^2)}}
\over x^2}~{{e^x-1+xe^x}\over {(e^x-1)^2}} \cr
&
=\int_a^{\infty}{dx\over x}e^{-x}(1+x)\bigl[1-{(a^2-b^2)\over 2x^2}+
\cdots\bigr] \cr
&
=e^{-a}\Bigl[1-{1\over 4}(1+a){(a^2-b^2)\over a^2}\Bigr]
-Ei(-a)\bigl[1+{(a^2-b^2)\over 4}\bigr]+\cdots,}}
where we have used
\eqn\ert{\int_a^{\infty}dx~{e^{-x}\over x^{n+1}}={(-1)^{n+1}\over n!}
Ei(-a)+{e^{-a}\over a^n}\sum_{m=0}^{n-1}{{(-1)^ma^m}\over{n(n-1)\cdots
(n-m)}},}
with
\eqn\erre{ Ei(-a)=-\int_a^{\infty}{dx\over x}e^{-x}.}
Since $Ei(-a)\to -e^{-a}/a$ as $a\to\infty$, we arrive at
\eqn\sse{\tilde I(a\to\infty,b)={e^{-a}\over a}\Bigl[1+a-{1\over 2}(a^2-b^2)
+\cdots\Bigr].}

The other integral shown in \efm\ has the following form:
\eqn\evj{ J(a,b)=\int_0^{\infty}
dz~{z\sqrt{z^2+b^2}\over\sqrt{z^2+a^2}}\Bigl(e^{\sqrt{z^2+a^2}}-1\Bigr)^{-1}.}
Before evaluating \evj\ fully, we first take the limit $b=0$ and write
\eqn\evj{ J(a)=J(a,0)=\int_0^{\infty}
dz~{z^2\over\sqrt{z^2+a^2}}\Bigl(e^{\sqrt{z^2+a^2}}-1\Bigr)^{-1}.}
Following the procedures outlined above, eq. \evj\ becomes
\eqn\evij{\eqalign{J(a)&={\pi^2\over 6}+\sum_{n=-\infty}^{\infty}\int dz~z^2
\Bigl[{1\over{z^2+a^2+4\pi^2n^2}}-{1\over{z^2+4\pi^2n^2}}\Bigr]
-{1\over 2}\int_0^{\infty}dz~\Bigl[{z^2\over\sqrt{z^2+a^2}}-z\Bigr] \cr
&
={\pi^2\over 6}+J^{(1)}(a)+J^{(2)}(a),}}
where we have added and subtracted
\eqn\evvij{\eqalign{ J(0)&=\int_0^{\infty}dz~{z\over{e^z-1}}
=\sum_{n=-\infty}^{\infty}\int dz~{z^2\over{z^2+4\pi^2n^2}}
-{1\over 2}\int_0^{\infty}dz~z \cr
&
={\pi^2\over 6}~.}}

Using
\eqn\iiy{ \intz{z^{2-\eps}\over{(z^2+\alpha_1^2)(z^2+\alpha_2^2)}}
={\pi\over2(\alpha_1^2-\alpha_2^2)}~{\rm sec}\bigl(
{\eps\pi\over 2}\bigr)\Bigl(\alpha_1^{1-\eps}-\alpha_2^{1-\eps}\Bigr),}
the first integral in \evij\ in its regularized form can be written as
\eqn\evyj{\eqalign{ J^{(1)}_{\eps}(a)&=-a^2\sumn\int_0^{\infty}dz~{z^{2-\eps}
\over{(z^2+4\pi^2n^2)(z^2+4\pi^2n^2+a^2)}} \cr
&
=-{\pi\over 2}~{\rm sec}\bigl(
{\eps\pi\over 2}\bigr)\sumn\Bigl[\bigl(4\pi^2n^2+a^2\bigr)^{(1-\eps)/2}
-(2\pi n)^{1-\eps}\Bigr] \cr
&
=-{\pi\over 2}~{\rm sec}\bigl({\eps\pi\over 2}\bigr)\Bigl\{
a^{1-\eps}+2~{\tilde J}^{(1)}_{\eps}(a)\Bigr\},}}
where
\eqn\evyii{{\tilde J}^{(1)}_{\eps}(a)= \sum_{n=1}^{\infty}\Bigl[\bigl(
4\pi^2n^2+a^2\bigr)^{(1-\eps)/2}-(2\pi n)^{1-\eps}\Bigr].}
Since
\eqn\devi{\eqalign{ {\partial{\tilde J^{(1)}_{\eps}(a)}\over\partial a}&
=a(1-\eps)\sum_{n=1}^{\infty}\bigl(4\pi^2n^2+a^2\bigr)^{-(1+\eps)/2}\cr
&
=a(1-\eps)\Biggl\{{\zeta(1+\eps)\over (2\pi)^{1+\eps}}+\sum_{n=1}^{\infty}
{1\over(2\pi n)^{1+\eps}}\Bigl[\Bigl(1+{a^2\over 4\pi^2n^2}\Bigr)^{-(1+\eps)/2}
-1\Bigr]\Biggr\} \cr
&
=a(1-\eps)(2\pi)^{-(1+\eps)}\zeta(1+\eps)+O(a^3),}}
this implies
\eqn\evyy{ {\tilde J}^{(1)}_{\eps}(a)=
{a^2\over 2}(1-\eps)(2\pi)^{-(1+\eps)}\zeta(1+\eps)+\cdots,}
or, in the vanishing $\eps$ limit,
\eqn\evye{ J^{(1)}_{\eps}(a)=-{{\pi a}\over 2}-{a^2\over 4}\Bigl[
{1\over\eps}-{\rm ln}{2\pi}+\gamma-1\Bigr]+\cdots .}

In a similar manner, we have
\eqn\evyye{ J^{(2)}(a)=-{1\over 2}\intz\Bigl[{z^2\over\sqrt{z^2
+a^2}}-z\Bigr]={a^2\over 4}\Bigl[{1\over\eps}-{\rm ln}
{a\over 2}\Bigr],}
with the help of
\eqn\eij{\int_0^{\infty}{dz~z^{2-\eps}\over\sqrt{z^2+a^2}}={a^{2-\eps}\over
2\sqrt{\pi}}~\Gamma({{3-\eps}\over 2})~\Gamma(-1+{\eps\over 2})={1\over\eps}
-{\rm ln}{a\over 2}+O(\eps),}
and discarding the $a$-independent term.
Adding up \evye\ and \evyye\ then leads the final result
\eqn\evvy{\eqalign{ J(a)&=\int_0^{\infty}{dz~z^2\over\sqrt{z^2+a^2}}\Bigl(
e^{\sqrt{z^2+a^2}}-1\Bigr)^{-1} \cr
&
={\pi^2\over 6}-{{\pi a}\over 2}-{a^2\over 4}\Bigl[~{\rm ln}{a\over 4\pi}
+\gamma-1 \Bigr]+\cdots.}}

Finally, to evaluate \evj, we observe that
\eqn\ers{ {{\partial J(a,b)}\over{\partial b}}=b~I(a,b).}
Integrating over $b$ with the help of \evy\ and imposing the boundary
condition \evvy\ gives
\eqn\evvyj{\eqalign{J(a,b)&={\pi^2\over 6}-{\pi\over 2}\sqrt{a^2-b^2}
-{1\over 4}\bigl(a^2-b^2\bigr)\Bigl[{\rm ln}{a\over{4\pi}}+\gamma
+{\rm tanh}^{-1}{b\over a}\Bigr] \cr
&
+\sqrt{a^2-b^2}~{\rm sin}^{-1}{b\over a}+{a^2\over 4}-b-{b^3\over 36}
+\cdots.}}
The above expression can be checked by taking the limit $a=b$ where
\evj\ can be simplified using \ref\arfken:
\eqn\debyef{ \int_0^r{du~u^{\ell}\over {e^u-1}}=r^{\ell}\Bigl[{1\over \ell}
-{r\over 2(\ell+1)}+\sum_{n=1}^{\infty}{{B_{2n}r^{2n}}\over
{(2n+\ell)(2n)!}}\Bigr] \qquad (\ell \ge 1),}
and leads to
\eqn\efr{\eqalign{ J^{*}(a)&=\int_0^{\infty}dz~{z\over{e^{\sqrt{z^2+a^2}}-1}}
=\int_a^{\infty}du~{u\over{e^u-1}} \cr
&
={\pi^2\over 6}-a\Bigl[1-{a\over 4}+\sum_{n=1}^{\infty}{B_{2n}a^{2n}\over
{(2n+1)!}}\Bigr].}}
With $B_2=1/6$, ones finds an amazing agreement between $J^{*}(a)$
and $J(a,b=a)$ in \evvyj. Similarly, for $b\to a\to\infty$, we have
\eqn\ervi{ J(a,b\to\infty)=e^{-a}\Bigl[1+{a\over 2}+{b^2\over 2a}
+\cdots\Bigr].}
%

\goodbreak
\bigskip
\centerline{\bf REFERENCES}
\medskip
\nobreak
\par\hang\noindent{\linde} A. Linde, {\it Rep. Prog. Phys.} Vol. {\bf 42}
(1979) 379;
D. A. Kirzhnits and A. D. Linde, {\it Phys. Lett.} {\bf 42B} (1972) 471.
\medskip
\par\hang\noindent{\muller} see for example, R. C. Hwa, ed.
{\it Quark Gluon Plasma}, (World Scientific, 1990); B. M\"uller,
{\it The Physics of the Quark-Gluon Plasma} (Springer-Verlag, 1985).
\medskip
\par\hang\noindent{\weinberg} S. Weinberg, {\it Phys. Rev.}
{\bf D9} (1974) 3357.
\medskip
\par\hang\noindent{\jackiw} L. Dolan and R. Jackiw, {\it Phys. Rev.}
{\bf D9} (1974) 3320.
\medskip
\par\hang\noindent{\bp} E. Braaten and R. D. Pisarski, {\it Phys. Rev. Lett.}
{\bf 64} (1990) 1338; {\it Nucl. Phys.} {\bf B337} (1990) 569.
\medskip
\par\hang\noindent{\elmfors} P. Elmfors, {\it Z. Phys.} {\bf C56} (1992) 601;
{\it Int. J. Mod. Phys.} {\bf A8} (1993) 1887; H. Matsumoto, Y. Nakano and
H. Umezawa, {\it Phys. Rev.} {\bf D29} (1984) 1116.
\medskip
\par\hang\noindent{\sypi} G. Amelino-Camelia and So-Young Pi, {\it Phys.
Rev.} {\bf D 47} (1993) 2356; MIT-CTP-2255, 1993.
\medskip
\par\hang\noindent{\oconnor} D. O'Connor and C. R. Stephens,
{\it Int. J. Mod. Phys.} {\bf A9} (1994) 2805, and references therein.
\medskip
\par\hang\noindent{\wils} K. Wilson, {\it Phys. Rev.} {\bf B4} (1971) 3174;
K. Wilson and J. Kogut, {\it Phys. Rep.} {\bf 12C} (1975) 75.
\medskip
\par\hang\noindent{\wetterich} N. Tetradis and C. Wetterich,
{\it Int. J. Mod. Phys. } {\bf A9} (1994) 4029;
{\it Nucl. Phys.} {\bf B422} (1994) 541 and {\bf B398} (1993) 659,
and references therein.
\medskip
\par\hang\noindent{\lpt} S.-B. Liao and J. Polonyi, {\it Nucl. Phys.}
{\bf A570} (1994) 203c; S.-B. Liao, J. Polonyi and D. P. Xu, Duke-TH-94-66,
to appear in {\it Phys. Rev. D}.
\medskip
\par\hang\noindent{\llp} S.-B. Liao and J. Polonyi, DUKE-TH-94-64 \& LPT
94-3, to appear in {\it Phys. Rev. D}.
\medskip
\par\hang\noindent{\coleman} S. Coleman and E. Weinberg, {\it Phys. Rev.}
{\bf D7} (1973) 1888.
\medskip
\par\hang\noindent{\lp} S.-B. Liao and J. Polonyi,
{\it Ann. Phys.} {\bf 222} (1993) 122.
\medskip
\par\hang\noindent{\landsman} E. L. M. Koopman and N. P. Landsman,
{\it Phys. Lett.} {\bf B223} (1989) 421; N. P. Landsman, {\it Nucl. Phys.}
{\bf B322} (1989) 498; D. Gross, R. Pisarski and A. Yaffe,
{\it Rev. Mod. Phys.} {\bf 53} (1981) 43.
\medskip
\par\hang\noindent{\babu} K. Babu Joseph, V. C. Kuriakose and M. Sabir,
{\it Phys. Lett.} {\bf B115} (1982) 120; P. Fendley, {\it Phys. Lett.}
{\bf B196} (1987) 175; H. A. Weldon, {\it Phys. Lett.} {\bf B174} (1986) 427;
K. Funakubo and M. Sakamoto, {\it Phys. Lett.} {\bf B186} (1987) 205.
\medskip
\par\hang\noindent{\gap} S.-P. Chia, {\it Int. J. Mod. Phys.} {\bf A2}
(1987) 713, and references therein.
\medskip
\par\hang\noindent{\dyson} O. J. P. Eboli and G. C. Marques,
{\it Phys. Lett.} {\bf B162} (1985) 169.
\medskip
\par\hang\noindent{\kuzmin} F. R. Klinkhammer and N. S. Manton, {\it Phys.
Rev.} {\bf D30} (1984) 2212; V. A. Kuzmin, V. A. Rubakov and M. E.
Shaposhnikov, {\it Phys. Lett.} {\bf B155} (1985) 36.
\medskip
\par\hang\noindent{\gauge} S.-B. Liao and J. Polonyi, DUKE-TH-94-65.
\medskip
\par\hang\noindent{\arfken} G. Arfken, {\it Mathematical Methods for
Physicists}, (Academic, New York, 1985), p. 338.
\medskip

\bigskip
\medskip
\ifnum\dofig=0
\centerline{\bf FIGURE CAPTIONS}
\medskip
\nobreak
\par\hang\noindent Fig. 1. Diagrammatic representation of loop resummation.
Contributions from all non-overlapping graphs are included in our physical
loop.
\medskip
\par\hang\noindent Fig. 2. Flow pattern of the blocked potential for
$T=0$ with $\tilde\mu_R^2=10^{-4}$, $\tilde\lambda_R=0.1$ and $\Lambda=10$.
Solid and dashed lines represent the RG and IMA results, respectively.
\medskip
\par\hang\noindent Fig. 3. Flow pattern of the blocked potential for
$T=5$ with $\tilde\mu_R^2=10^{-4}$, $\tilde\lambda_R=0.1$ and $\Lambda=10$.
Solid and dashed lines represent the RG and IMA results, respectively.
\medskip
\par\hang\noindent Fig. 4. Evolution of $\mu_{\beta,k}^2$ and
$\lambda_{\beta,k}$ as a function of $k$ for various values of $T$ using
$\tilde\mu_R^2=10^{-4}$, $\tilde\lambda_R=0.1$ and $\Lambda =20$.
\medskip
\par\hang\noindent Fig. 5. Comparison of the flow of the mass parameter
at $T=0$ for different RG prescriptions using
$\tilde\mu_R^2=10^{-4}$, $\tilde\lambda_R=0.1$, $\tilde{\Lambda}=30$, and
$\Lambda = \tilde\Lambda/\sqrt{2}$.
\medskip
\par\hang\noindent Fig. 6. Comparison of the flow of the coupling constant
at $T=0$ for different RG prescriptions using
$\tilde\mu_R^2=10^{-4}$, $\tilde\lambda_R=0.1$, $\tilde{\Lambda}=30$, and
$\Lambda = \tilde\Lambda/\sqrt{2}$.
\medskip
\par\hang\noindent Fig. 7. Limitation on dimensional reduction at
high temperature. Notice the gap between the mass parameters generated from the
dimensionally reduced and the full RG prescriptions.
\medskip
\par\hang\noindent Fig. 8. Temperature dependence for $\mu_{\beta}^2$
and $\lambda_{\beta}$ at $k=0$.
\medskip
\par\hang\noindent Fig. 9. Comparison of the critical temperature $T_c$
obtained
by RG with $\tilde T_c$. Notice that a higher value is predicted for RG.
\medskip
\par\hang\noindent Fig. 10. Temperature dependence of various quantities in
the symmetry broken phase.
\fi
\end